\documentclass[fleqn,usenatbib]{mnras}

\usepackage[T1]{fontenc}

\DeclareRobustCommand{\VAN}[3]{#2}
\let\VANthebibliography\thebibliography
\def\thebibliography{\DeclareRobustCommand{\VAN}[3]{##3}\VANthebibliography}

\usepackage{graphicx}	
\usepackage{amsmath}	
\usepackage{amssymb}	

\newcommand{\Rwd}{\mbox{$R_{\mathrm{wd}}$}}
\newcommand{\Mwd}{\mbox{$M_{\mathrm{wd}}$}}
\newcommand{\Msun}{\mbox{$\mathrm{M}_{\odot}$}}
\newcommand{\Rsun}{\mbox{$\mathrm{R}_{\odot}$}}

\newcommand{\bs}{\ensuremath{\langle \vert B \vert \rangle}}

\newcommand{\vsini}{\ensuremath{v\,\sin\,i}}
\newcommand{\kms}{km\,s$^{-1}$}

\newcommand{\Ion}[2]{#1{\,\textsc{#2}}}
\newcommand{\Teff}{\mbox{$T_{\mathrm{eff}}$}}
\newcommand{\logg}{\mbox{$\log g$}}

\usepackage{newtxtext,newtxmath}

\title[A pre-intermediate polar]{Discovery of a young pre-intermediate polar}

\author[Wilson et al.]{David J. Wilson$^{1}$\thanks{djwilson394@gmail.com}, Odette Toloza$^{2,3}$, John D. Landstreet$^{4,5}$,
Boris T. G{\"a}nsicke$^{2}$, \newauthor Jeremy J.  Drake$^{5}$, J. J. Hermes$^{7}$, Detlev Koester$^{8}$  \medskip\\
$^{1}$ McDonald Observatory, University of Texas at Austin, 2515 Speedway, C1402, Austin, TX 78712, USA \\ 
$^{2}$ Department of Physics, University of Warwick, Coventry CV4 7AL, UK \\
$^{3}$ Departamento de Física, Universidad Técnica Federico Santa María, Avenida España 1680, Valparaíso, Chile\\
$^{4}$ Armagh Observatory \& Planetarium, Armagh, BT61 9DG, Northern Ireland, and  \\
$^{5}$ Department of Physics \& Astronomy, University of Western Ontario, London, ON N6G 1P7 \\
$^{6}$ Center for Astrophysics | Harvard \& Smithsonian, 60 Garden Street, Cambridge, MA 02138, USA\\
$^{7}$ Department of Astronomy, Boston University, 725 Commonwealth Ave., Boston, MA 02215, USA\\
$^{8}$Institut f{\"u}r Theoretische Physik und Astrophysik, University of Kiel, 24098 Kiel, Germany 
}

\date{Accepted 2021 August 25. Received 2021 August 23; in original form 2021 June 30}

\pubyear{2021}

\begin{document}
\label{firstpage}
\pagerange{\pageref{firstpage}--\pageref{lastpage}}
\maketitle

\begin{abstract}
We present the discovery of a magnetic field on the white dwarf component in the detached post common envelope binary (PCEB) CC\,Cet. Magnetic white dwarfs in detached PCEBs are extremely rare, in contrast to the high incidence of magnetism in single white dwarfs and cataclysmic variables. We find Zeeman-split absorption lines in both ultraviolet \textit{Hubble Space Telescope} (\textit{HST}) spectra and archival optical spectra of CC\,Cet. Model fits to the lines return a mean magnetic field strength of  \bs\ $\approx$ 600--700\,kG. Differences in the best-fit magnetic field strength between two separate \textit{HST} observations and the high \vsini\ of the lines indicate that the white dwarf is  rotating with a period $\sim0.5$\,hours, and that the magnetic field is not axisymmetric about the spin axis. The magnetic field strength and rotation period are consistent with those observed among the intermediate polar class of cataclysmic variable, and we compute stellar evolution models that predict CC\,Cet will evolve into an intermediate polar in 7--17\,Gyr. Among the small number of known PCEBs containing a confirmed magnetic white dwarf, CC\,Cet is the hottest (and thus youngest), with the weakest field strength, and cannot have formed via the recently proposed crystallisation/spin-up scenario. In addition to the magnetic field measurements, we update the atmospheric parameters of the CC\,Cet white dwarf via model spectra fits to the \textit{HST} data and provide a refined orbital period and ephemeris from \textit{TESS} photometry.       
\end{abstract}




\section{Introduction}
Post Common Envelope Binaries (PCEBs) are systems containing at least one evolved star, in which the initial separation was close enough that the secondary was engulfed by the expanding envelope of the primary as it passed through the giant stages of its evolution. After double-degenerates, the second most common type of PCEBs are white dwarfs plus main-sequence companions \citep{toonenetal17-1}. The vast majority of the known systems of this kind contain a white dwarf and an M~dwarf companion \citep{rebassa-mansergasetal10-1}, although this is a selection effect as identification requires the white dwarf to be detectable against its main-sequence companion \citep{inightetal21-1}. For the remainder of this paper, we will use PCEB as synonymous for a close, detached binary consisting of a white dwarf and a main sequence companion that formed through a common envelope. The common envelope dramatically shrinks the binary separation, leaving most PCEBs with orbital periods of $\approx$\,0.1--5\,d \citep{nebotetal11-1}. After emerging from the common envelope, PCEBs lose angular momentum via gravitational radiation \citep{paczynski+sienkiewicz81-1}, and, if the main-sequence component possesses a convective envelope, magnetic wind braking \citep{rappaportetal83-1}. Consequently, the orbital separation decreases, eventually bringing the system into a semi-detached configuration, starting Roche-lobe overflow mass transfer from the companion onto the white dwarf~--~at this point, the PCEB will have evolved into a cataclysmic variable (CV). 

Because these are stages along an evolutionary path, fundamental physical characteristics that are not expected to be affected by age or the mass transfer process should have the same distributions in both the PCEB and CV populations. This turns out not to be the case. Most prominently, the occurrence rate of white dwarfs with detectable magnetic fields is hugely discrepant between the two populations \citep{liebertetal05-2}. 

In a volume-limited sample of 42 CVs, \citet{palaetal20-1} found that $36\pm7$\,per\,cent of the white dwarf primaries had magnetic field strengths $\gtrsim 1$\,MG. Magnetic CVs are divided into polars with $B \gtrsim 10$\,MG, where material from the secondary is accreted directly onto the poles of the white dwarf along the magnetic field lines, and intermediate polars with  $0.1 \la B \lesssim 10$\,MG, in which an accretion disc is usually formed but truncated at the magnetospheric radius. Whereas the white dwarf spin periods of polars are locked to the binary period, the periods of intermediate polars are highly asynchronous, with some exhibiting spin periods of just a few tens of seconds \citep{patterson79-1,lopesdeoliveiraetal20-1}.

The lack of an accretion disc around polars allows the white dwarf photosphere to be detected, enabling robust characterisation of the white dwarf parameters via spectral fitting and of the magnetic field strength and structure via detection of Zeeman-split lines and/or cyclotron emission \citep[e.g.][]{schwope90-1, gaensickeetal04-1, ferrarioetal95-1}. Conversely, very few robust measurements of the characteristics and magnetic field strengths of white dwarfs in intermediate polars exist, as the white dwarf is typically outshone by the accretion flow and the fields are too low to be detected via cyclotron emission from the accretion regions on the white dwarf. Variable polarised emission has been detected in a few intermediate polars \citep[e.g. ][]{potter+buckley18-1}, providing loose constraints on their magnetic fields strengths consistent with the expected $B \lesssim 10$\,MG range. Detecting and characterising ``Pre-intermediate polars'', that is, white dwarfs with magnetic fields of  $B \lesssim 10$\,MG in detached binaries that have yet to be begin mass transfer, would therefore provide a useful insight into the unobservable properties of the intermediate polar population, such as the distributions of magnetic field strength and white dwarf mass. 

However, in stark contrast to the CVs, magnetic white dwarfs in detached PCEBs are extremely rare. In a spectroscopic survey of over 1200 detached white dwarf plus M~dwarf binaries, \citet{silvestrietal07-1} found just two candidate magnetic white dwarfs (which to our knowledge have not yet been confirmed or refuted), while \citet{liebertetal15-1} found no magnetic white dwarfs in a sample of 1735 binaries. 

There are currently 16 PCEBs known with fields $ > 10$\,MG \citep{reimersetal99-1, reimers+hagen00-1, schmidtetal05-1, schmidtetal07-1, schwopeetal09-1, parsonsetal21-1}, which were identified via of the detection of either cyclotron or Balmer emission lines, spectral features that may have led to them being inadvertently excluded from the samples mentioned above. In the intermediate polar range, SDSS\,J030308.35+005444.1 \citep[$B = 8$\,MG,][]{parsonsetal13-1} is the only unambiguous detection\footnote{A second, ambiguous case is the prototype PCEB V471\,Tauri: A low signal to noise ratio (S/N) feature consistent with Zeeman splitting of the \ion{Si}{iii}\,1207.5\,\AA\ absorption line by a $\sim 350$\,kG field was found by \citet{sionetal98-2}, but extensive spectroscopic follow-up failed to detect Zeeman splitting at any other lines \citep{sionetal12-1}.}. These systems were initially thought to simply be polars with very low accretion rates \citep[``LARPs'', e.g. ][]{reimersetal99-1}, but it was soon realised that the companion stars were not quite filling their Roche lobes \citep[e.g. ][]{vogeletal07-1} and the observed accretion was from stellar wind rather than mass transfer. They were therefore classified as pre-polars or PREPs, and we refer to them by that term henceforth. However, as discussed by \citet{parsonsetal21-1} and \citet{schreiberetal21-1},  it is likely that these systems were non-magnetic CVs in the past, which formed their magnetic fields via a combination of mass transfer induced spin up and crystallisation, entering a detached phase in the process. The secondaries are predicted to refill their Roche lobes in the future and thus the term pre-polar remains fitting, but these systems do not represent examples of the ``missing'' systems identified by \citet{liebertetal05-2}; that is, magnetic white dwarfs in young PCEBs that have yet to undergo a period of mass transfer.         

Here we present the first unambiguous detection of a magnetic white dwarf in a young PCEB. CC\,Cet (PG 0308+096) was identified as a post common envelope binary by \citet{safferetal93-1} via radial velocity measurements of the H$_{\alpha}$ line, with a $\approx 6.9$\,h orbital period confirmed by \citet{somersetal96-2}. The system consists of a low-mass white dwarf ($\approx 0.4$\,\Msun) with effective temperature (\Teff)\,$\approx 25000$\,K and an M4.5--5\, dwarf secondary \citep{tappertetal07-2} at a distance of $121.4\pm 1.1$~pc \citep{gaia18-1}. We have obtained ultraviolet spectroscopy of the white dwarf component, revealing Zeeman-split absorption lines induced by a 600--700\,kG magnetic field, indicating that CC\,Cet is a young, low-mass pre-intermediate polar. 

The paper is arranged as follows: Section \ref{sec:obs} describes the observations of CC\,Cet; Section \ref{sec:models} presents our model fitting process to the spectra to measure the white dwarf atmospheric parameters and magnetic field characteristics; in Section \ref{sec:dis} we model the evolution of CC\,Cet and discuss the implications of our results for the study of magnetic white dwarfs in binaries. We conclude in Section \ref{sec:conc}.  


\section{Observations}
\label{sec:obs}
Details of our spectroscopic observations of CC\,Cet are collated in Table \ref{tab:hst_obs}.

\subsection{\textit{Hubble Space Telescope}}

\begin{figure*}
    \centering
    \includegraphics[width=2\columnwidth]{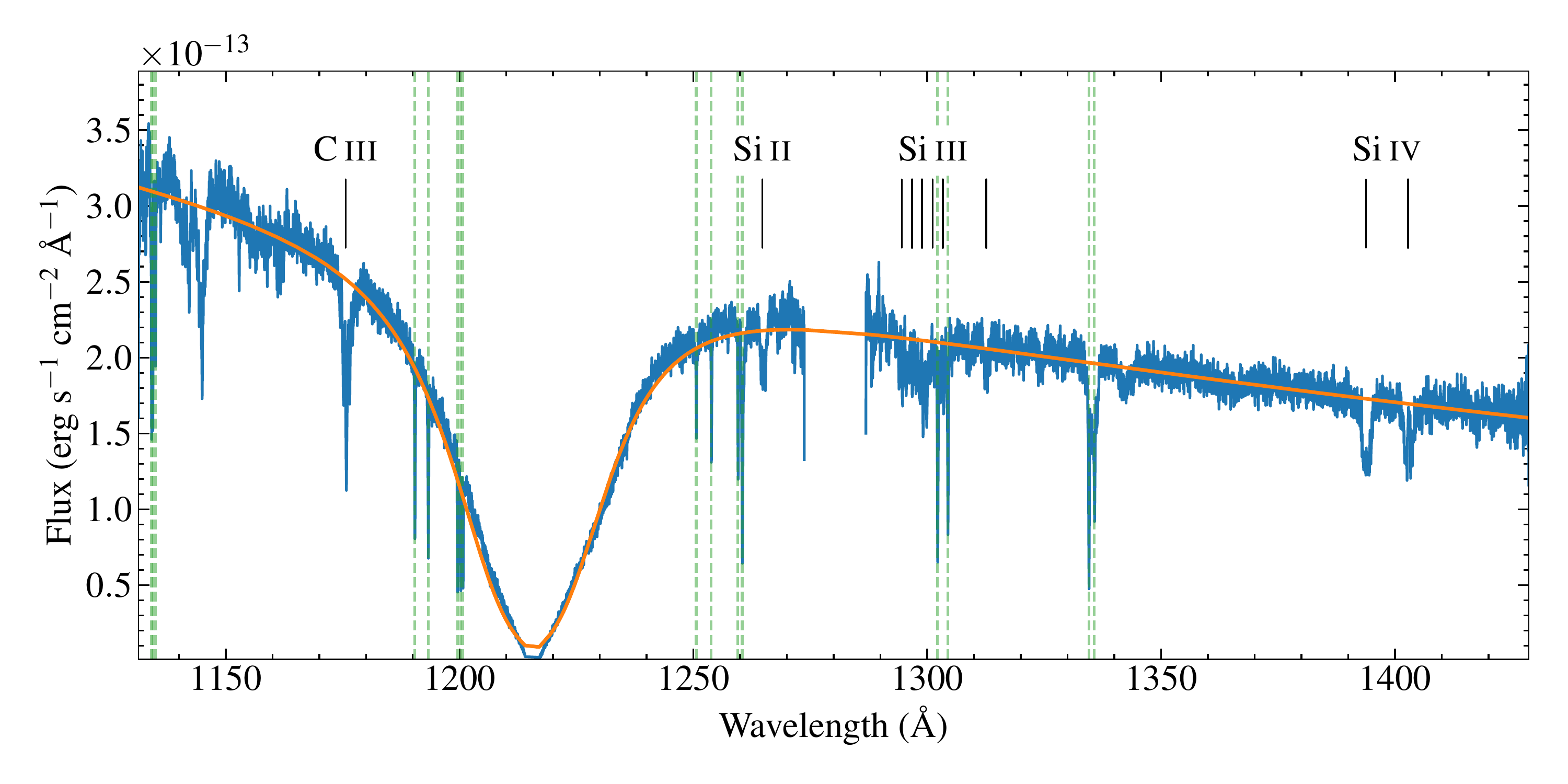}
    \caption{Full COS G130M spectrum of CC\,Cet obtained on 2018~February~01. The spectrum has been smoothed by a 5-point boxcar, matching the approximate oversampling rate of the COS detector. The best-fit model atmosphere spectrum is overplotted in orange. Rest wavelengths of the C and Si absorption lines discussed in the text are labeled, and interstellar absorption lines are marked with green dashed lines.}
    \label{fig:cos_spec}
\end{figure*}

CC\,Cet was observed with the Cosmic Origins Spectrograph \citep[COS,][]{greenetal12-1} onboard the \textit{Hubble Space Telescope} (\textit{HST}) as part of program ID 15189 (PI Wilson). Observations were obtained on 2018~February~1 and 2018~July~22 with exposure times of 1865\,s each, using the G130M grating with a central wavelength of 1291\,\AA. The spectral resolution of this setup is $\approx 0.1$\,\AA\ per resolution element, or $\approx20$\,\kms. We abbreviate the two visits as the 2018~Feb and 2018~Jul spectra respectively throughout. The spectra were reduced using the standard \textsc{calcos} tools. Figure \ref{fig:cos_spec} shows the 2018~Feb spectrum with the absorption lines discussed below marked. Due to the combination of the radial velocity shifts induced by the binary orbit (see Section \ref{sec:rotation}) and the magnetic splitting discussed below, we did not attempt to coadd the two spectra, instead analysing each one separately. 

Of the other targets observed in program 15189, we use LM\,Com as a comparison example of a non-magnetic white dwarf with a similar \Teff\ and surface gravity (\logg) to CC\,Cet in several plots below. LM\,Com was observed on 2017~December~17 with an exposure time of 1815\,s and otherwise the same details as the CC\,Cet observations.

\subsection{\textit{TESS}}
\defcitealias{lightkurve18-1}{Lightkurve Collaboration, 2018}

\begin{figure}
    \centering
    \includegraphics[width=\columnwidth]{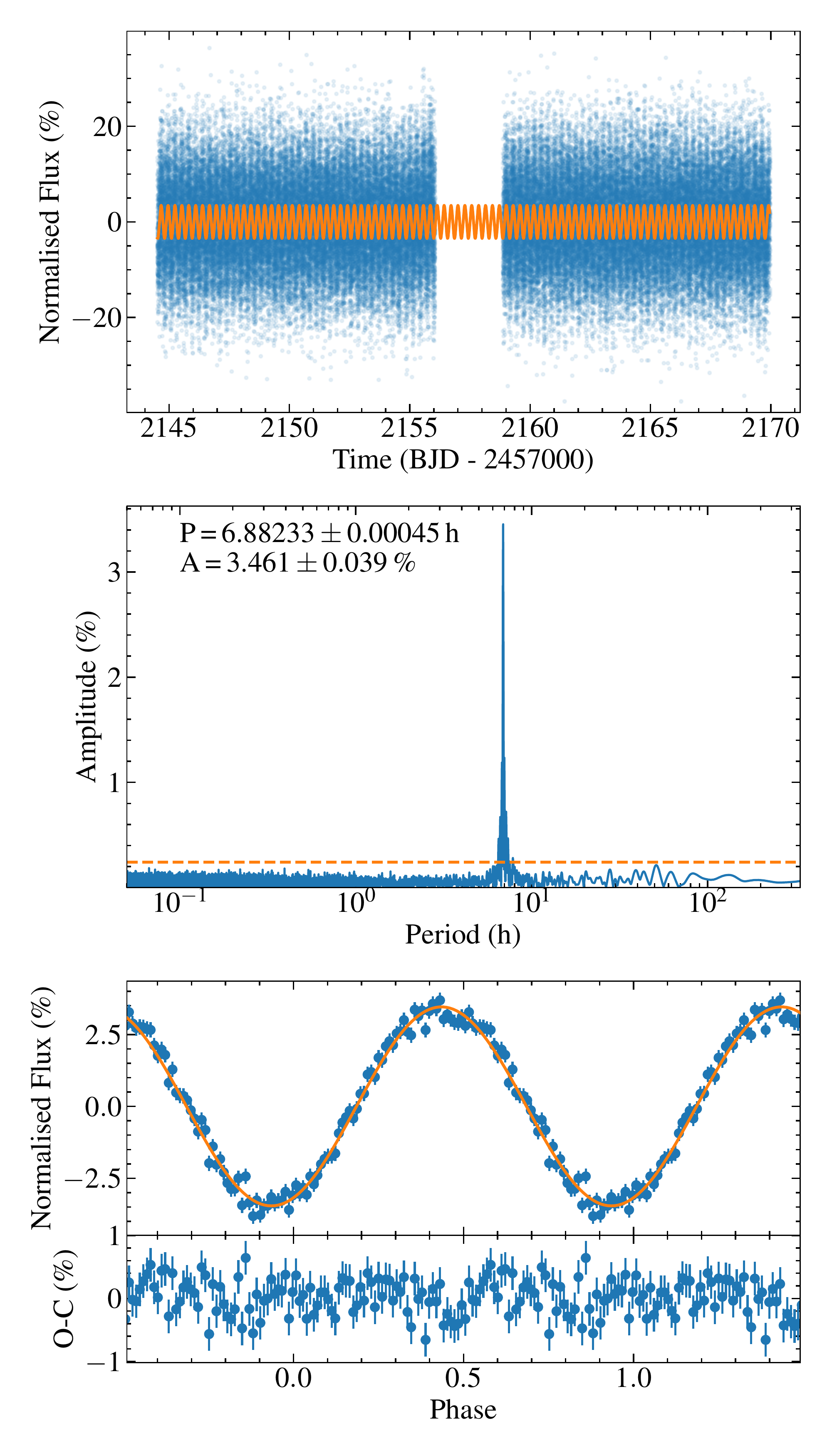}
    \caption{Top panel: \textit{TESS} photometry of CC\,Cet with 20\,s cadence (blue), with a model fit in orange. Middle panel: Lomb-Scargle periodogram of the \textit{TESS} light curve, clearly showing the binary orbital period but with no evidence for any other periodicities. The orange dashed line shows the 0.01\% false alarm probability. Bottom panel: Phase-folded light curve (blue) and sine model fit (orange), repeated for clarity. The residuals are shown as an O-C calculation.}  
    \label{fig:tess}
\end{figure}

CC\,Cet was observed by the \textit{Transiting Exoplanet Survey Satellite} (\textit{TESS}, TIC 337219837) in Sector 4 (2018~October~19--2018~November~14) at 30\,min cadence and again in Sector 31 (2020~October~12--2020~November~22), for which both  2\,min and 20\,s cadence data was returned (GO 3124, PI Hermes). Figure \ref{fig:tess} shows the results of our analysis of the 20\,s data using \textsc{Lightkurve} \citepalias{lightkurve18-1}. The false alarm probability was calculated following the recipe from \citet{belletal19-1}. The light curve shows clear sinusoidal variation with a period of $6.88233\pm0.00045$\,h, in agreement with the orbital period of the binary system measured by \citet{somersetal96-2}. The photometric modulation has a constant amplitude, confirming that it is produced by irradiation of the M\,dwarf, which varies on the orbital period, rather than ellipsoidal modulation, which would induce an asymmetric double-peaked light curve. We calculated an updated ephemeris of TJD\,$=2459157.14634\pm0.00052$, defined as the time of inferior conjunction nearest to the mid-point of Sector 31.  

The results from the 20\,s data were double-checked against a 30\,min cadence light curve extracted from the Sector 4 data using the \texttt{eleanor} package \citep{feinsteinetal19-1}, with no evidence found for significant changes in either the
period or the amplitude of the modulation. Inspecting the power spectrum of the 20\,s light curve, we find no evidence for additional periods between 40\,s and $\approx14$\,d, adopting the 0.01\,per\,cent False Alarm Probability of 0.25\,per\,cent amplitude as an upper limit. We therefore detect no evidence for flux modulations induced by the white dwarf rotation (see Section \ref{sec:rotation}) or any evidence that the M\,dwarf rotation is not tidally locked to the orbital period. Splitting the light curve into  0.5\,d chunks, we found that the amplitude remained stable to within 1\,$\sigma$ over the sector, and hence we do not detect evidence for modulation due to spots on the secondary star and/or differential rotation, as seen in V471\,Tauri by \citet{kovarietal21-1}. A visual inspection of the 20\,s light curve found no significant flare events.  

\subsection{\textit{XMM-Newton}}

We obtained X-ray observations of CC Cet with the {\em XMM-Newton} (\textit{XMM}) space telescope timed to overlap the two \textit{HST}/COS visits: on 2018 February 1 at 23:05:40~UTC for a nominal exposure time of 5127s; and on 2018 July 22 at 03:56:09~UTC for a nominal exposure time of 15423s. The EPIC pn data were obtained in Large Window mode using the thin filter, with MOS1 and MOS2 employing Partial Window mode and the Medium filter. Data were processed using standard extraction methods within the \textit{XMM} {\tt Science Analysis System} version 18.0.0 to extract images, light curves, and spectra.  

\begin{figure}
    \centering
    \includegraphics[width=\columnwidth]{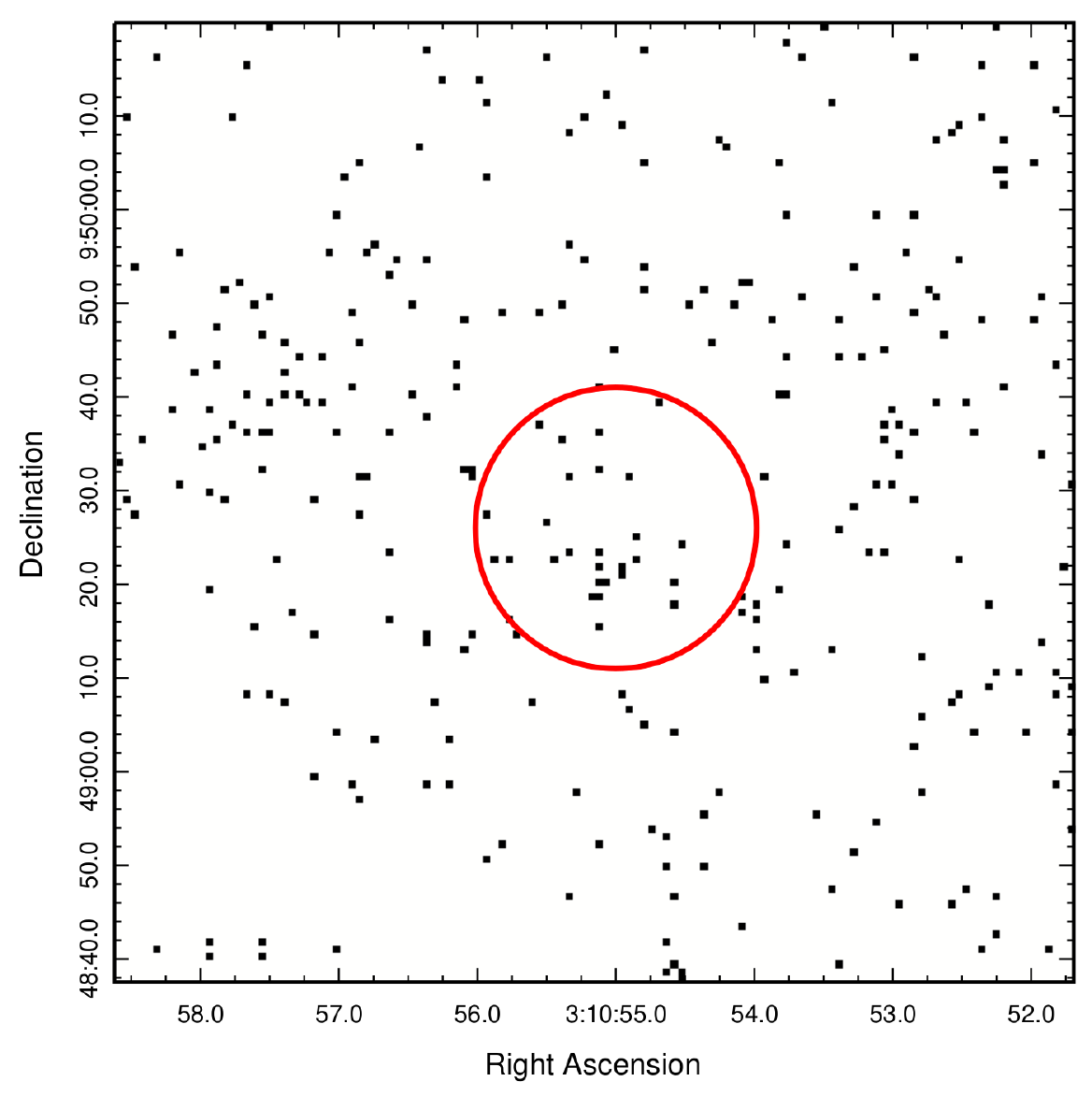}
    \caption{The {\it XMM-Newton} image in the pn detector of the vicinity around CC\,Cet. The red circle is centered on the position of CC\,Cet and has a radius of 15~arcsec. The 2x binned pixel scale is approximately 0.8 arcsec. The source is only detected at $<3\sigma$ significance, providing an upper limit on the X-ray flux.}  
    \label{f:pn}
\end{figure}

Unfortunately, both observations were severely afflicted with background flares, the first to such an extent that only a few hundred seconds of exposure were useful.  By-eye inspection of those data revealed no obvious signs of a strong source, and so in the following we ignore the scant remaining 2018 February data. For the second observation, 6011s and 7828s of useful exposure time were retrieved for pn and MOS detectors, respectively. The pn image in the vicinity of the source position is illustrated in Figure~\ref{f:pn}.

The MOS1 and MOS2 data were combined prior to analysis. Source counts were extracted from both pn and MOS data using a circular region with a radius of 15~arcsec enclosing approximately $70\pm 5$\%\ of the energy.  This size of region was chosen because the \textit{XMM} point spread function has very extended wings and enlarging the extraction region to encircle a greater fraction of the energy comes at the expense of significantly increasing the background.  An annular region of area greater than ten times the source region and centered on the source position was used for estimating the background rate.  The resulting count rates are listed in Table~\ref{t:xmm}. While the source appears at a level greater than the background in both pn and MOS, neither represents a significant detection at greater than the $3\sigma$ level.   

\begin{table} 
\setlength{\tabcolsep}{4pt} 
\caption{Summary of {\it XMM-Newton} results}
\begin{tabular}{lcc}
\hline
Parameter & pn & MOS\\
\hline 
Net exposure (s) & 6011 & 7828 \\
$15\arcsec$ radius count rate  (count~ks$^{-1}$) & $4.99\pm 0.91$ & $3.45\pm 0.66$ \\
Scaled background (count~ks$^{-1}$) & $3.17\pm 0.18$ &  $1.87\pm 0.13$\\
Net Source rate (count~ks$^{-1}$) & $1.82\pm 0.93$ & $1.58\pm 0.66$ \\
$L_X$ at $10^7$~K $(10^{27}$ erg s$^{-1}$)$^a$ & $6.1\pm 2.3$& $10.8\pm 4.6$\\
\hline
\end{tabular} 
\label{t:xmm} \\
{\footnotesize $^a$X-ray luminosity assuming an optically-thin plasma radiative loss model with solar metallicity and an interstellar absorbing column of $2\times 10^{20}$~cm$^{-2}$ (see text).}
\end{table} 

\subsection{VLT/UVES}
CC\,Cet was observed with the Ultraviolet and Visual Echelle Spectrograph \citep[UVES,][]{Dekkeretal00-1} on the Very Large Telscope (VLT) as part of the Supernovae Type Ia Progenitor Survey  \citep[SPY,][]{napiwotzkietal20-1, koesteretal09-2}. Two spectra were obtained on 2001~February~7--8, each covering the wavelength range 3281--6686\,\AA\ with $R\approx21\,000$. We retrieved both spectra as fully calibrated data products from the ESO Archive Science Portal\footnote{\url{http://archive.eso.org/scienceportal/home}}.

\section{Modelling}
\label{sec:models}
\subsection{White dwarf characteristics}
\label{sec:wdparams}
Figure \ref{fig:cos_spec} shows the 2018~Feb COS G130M spectrum of CC\,Cet. The spectrum is typical of white dwarfs in this temperature range \citep{koesteretal14-1}: dominated by the broad \ion{H}{i}\,1215.67\,\AA\ Lyman $\alpha$ line, along with a mixture of narrow and deep interstellar lines and broader, shallower photospheric lines. We detect no contribution from emission lines produced by the M\,dwarf, which is unsurprising given that the lack of flares in the \textit{TESS} light curve indicates that the M\,dwarf is relatively inactive. The spectrum shows photospheric absorption lines of \Ion{Si}{iv}, implying that the \Teff\ of the white dwarf is higher than 25\,000\,K.

To estimate the atmospheric parameters of the white dwarf in CC\,Cet, i.e. \Teff\ and \logg, we fitted the continuum of the COS spectroscopy with a grid of synthetic white dwarf models computed with an updated version of the code described in \citet{koester10-1}, using the Markov Chain Monte Carlo (MCMC) technique. The Eddington flux ($f_\text{Edd}$) of the models were scaled as
\begin{equation}
    F_{\rm obs} = 4\,\pi\,\left(\Rwd \times \Pi\right)^{2}\,\times\,f_{\rm Edd}(\Teff, \logg),
    \label{eq:model}
\end{equation}
where $\Pi$ is the parallax and \Rwd\ is the white dwarf radius. \Rwd\ is a function of \logg\ and \Teff\ via the white dwarf mass-radius relation. We used the mass-radius relation for white dwarfs with hydrogen-rich atmospheres by interpolating the cooling models from \citet{fontaineetal01-1} with thick hydrogen layers of $M_{\rm H}/\Mwd=10^{-4}$, which are available from the University of Montreal website\footnote{\href{http://www.astro.umontreal.ca/~bergeron/CoolingModels}{http://www.astro.umontreal.ca/$\sim$bergeron/CoolingModels}, \citet{bergeronetal95-2, holberg+bergeron06-1, tremblayetal11-2,kowalski+saumon06-1}.} 
In addition the models were corrected by reddening ($E(B-V)$) using the extinction parameterization from  \citet{fitzpatrick99-1}. In summary, the parameters to be fitted are: \Teff, \logg, $\Pi$, and $E(B-V)$.

We set a flat prior on $\Pi$ using the \textit{Gaia} DR2, parallax for CC\,Cet \citep[$\Pi=8.2381\pm0.0758$\,mas, \textit{Gaia} source id~=~15207693216816512][]{gaia18-1}, which corresponds to a distance of $D=121.4\pm1.1$\,pc\footnote{Note this analysis was carried out before the release of \textit{Gaia}\, EDR3, but the improvement in astrometric accuracy from DR2 to EDR3 was small so will not significantly change the parameters given here.}, and forced the fits to find the best parallax value within $1\sigma$. We set a Gaussian prior on the reddening of $E(B-V)=0.012 \pm 0.015$\,mag based on the measurement from the STructuring by Inversion the Local Interstellar Medium (stilism)\footnote{\href{https://stilism.obspm.fr/}{https://stilism.obspm.fr/}, \citet{lallementetal14-1, lallementetal18-1, capitanioetal17-1}.} at a distance of 120\,pc. \Teff\ and \logg\ were constrained to the values covered by our grid of model spectra.

The models have a pure hydrogen atmosphere with the parameter for mixing length convection set to 0.8. The grid spans $\Teff=10\,000-35\,000$\,K in steps of 200\,K, and $\logg=7.0-9.0$\,dex in steps of 0.1\,dex.

Earth airglow emission lines from Lyman\,$\alpha$ and metal absorption lines from the interstellar medium (Table\,\ref{tab:ISlines}) and the white dwarf photosphere (\ion{C}{iii}, \ion{Si}{ii}, \ion{Si}{iii},\ion{Si}{iv}, see below), were masked out during the process (see Figure \ref{fig:cos_spec}).

We used the python-based \texttt{emcee} MCMC method \citep{foreman-mackeyetal13-1}, where 100 walkers were sampling the parameter space during 10\,000 iterations. The likelihood function was defined as -0.5$\chi^{2}$. In general the walkers converged quickly, therefore we removed the first 250 steps from the chain. The full results are shown in Figure \ref{fig:mcmc}. The samples of \Teff, \logg, and $E(B-V)$ follow a normal distribution, except the samples of the parallax which clustered towards the lower tail of the distribution. While the parallax was tightly constrained during the fits to be within $\pm1\sigma$ from the \textit{Gaia} average value, the results hint that larger distances are required to improve the far-ultraviolet spectroscopic fits of CC\,Cet. For the normal distribution, we considered the median as the best value and the 15.9$^{\rm th}$ and 84.1$^{\rm th}$ percentiles for one standard deviation as error. The intrinsic uncertainties from the MCMC method are very small and are purely statistical. The results are $\Teff=25\,245^{+18}_{-19}$\,K, $\logg=7.606^{+0.005}_{-0.004}$\,dex, $E(B-V) = 0.0183\pm0.0005$\,mag for the 2018~Feb COS spectrum and $\Teff=25\,162^{+19}_{-20}$\,K, $\logg=7.564\pm0.005$\,dex, and $E(B-V) =0.023\pm0.0005$\,mag for the 2018~Jul spectrum. We computed the mean and standard deviation using the two estimates of the parameters from the fits which account for systematic errors. These best-fit values are quoted in Table \ref{tab:characteristics}. We generated distributions for the mass and radius by drawing 10\,000 pairs from normal distributions of $\Teff=25\,203\pm42$\,K and $\logg=7.58\pm0.02$\,dex, then computing the mass and radius for each pair. We find two normal distributions for the mass and radius described by $\Mwd=0.441\pm0.008$\,\Msun\ and $\Rwd=0.0179\pm0.0003$\,\Rsun.

\begin{table} 
\setlength{\tabcolsep}{4pt} 
\centering 
\caption{Characteristics of the CC\,Cet system. 
References: 1. This work; 2. \citet{gaia18-1}; 3. \citet{safferetal93-1}; 4. \citet{tappertetal07-2}; 5. \citet{somersetal96-2}.}
\begin{tabular}{llr}
\hline
Parameter & Value & Reference \\
\hline 
$T_{\mathrm{eff}}$\,(K)           & $25\,203\pm42$              & 1\\
$\log g$\,(cm\,s$^{-2}$)          &  $7.58\pm0.02$            & 1\\
$E(B-V)$\,(mag)                   & $0.021\pm0.002$           & 1\\
White dwarf Mass (\Msun)          & $0.441\pm0.008$           & 1\\
White dwarf Radius (\Rsun)        & $0.0179\pm0.0003$         & 1\\
Parallax (mas)                    & $8.2381\pm0.0758$         & 2 \\
Distance (pc)                     & $121.4\pm1.1$             & 1\\
Magnetic field strength (kG)      & 600--700                  & 1 \\
$v \sin i$ (km\,s$^{-1}$)         & $40\pm10$                 & 1\\
Secondary Mass (\Msun)            & $0.18\pm0.05$             & 3\\
Secondary Spectral Type           & M4.5--5                   & 3, 4 \\ 
Orbital period (\textit{TESS}, h) & $6.88233\pm0.00045$        & 1\\
Ephemeris (\textit{TESS}, TJD)   & $2\,459\,157.14634\pm0.00052$ & 1 \\
Binary inclination ($^{\circ}$)   & $35\pm5.5$                & 5 \\

\hline
\end{tabular} 
\label{tab:characteristics} 
\end{table} 

\subsection{Spectral lines in the COS spectra of CC Cet}

The COS spectra of CC\,Cet contain multiple absorption lines of both interstellar and photospheric origin. The interstellar lines are invariably due to resonance lines of neutral or singly-ionised abundant elements such as C\,{\sc ii}, N\,{\sc i}, O\,{\sc i}, Si\,{\sc ii}, and S\,{\sc ii}. Because the observed interstellar lines arise only from the lowest ground state level, it can sometimes happen that in a multiplet of resonance lines which is also present in the photospheric spectrum, only some of the lines of the multiplet are contaminated by interstellar scattering. This allows the photospheric lines to be reliably modelled without contamination from the interstellar lines. 

The strongest photospheric lines present in the COS spectrum of CC\,Cet are primarily due to C\,{\sc iii} (six lines at 1175\,\AA), Si\,{\sc iii} (six lines between 1294 and 1303\,\AA), and Si\,{\sc iv} (two lines at 1393 and 1402\,\AA). There are also two blended photospheric lines of Si\,{\sc ii} at 1265\,\AA.

\subsection{Magnetic field}

\begin{figure}
    \centering
    \includegraphics[width=9 cm]{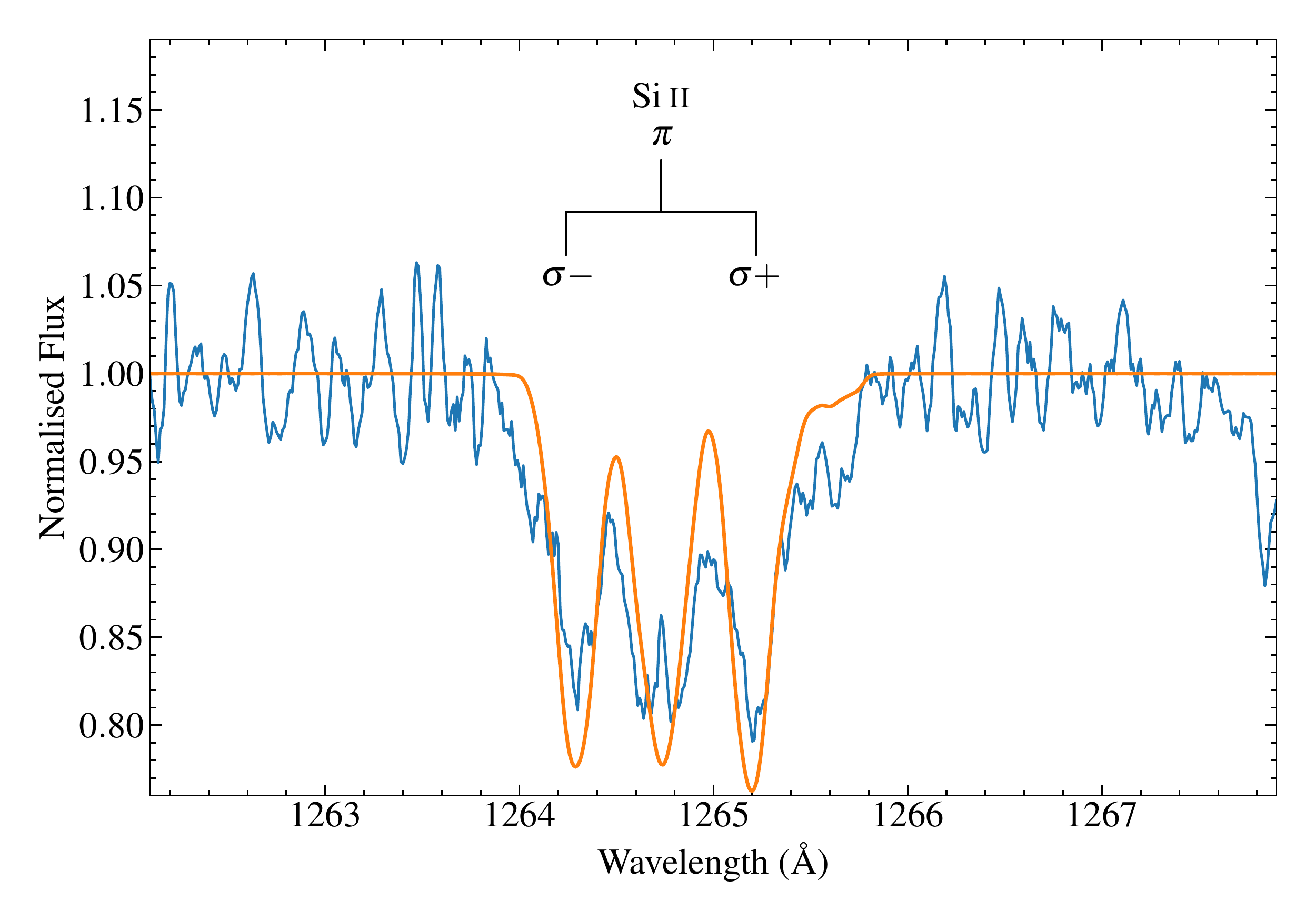}
    \caption{The \ion{Si}{ii} 1264.738\,\AA\ line in the 2018~Feb spectrum (blue) compared with a model fit (orange). The line is Zeeman-split into a central $\pi$ and outer $\sigma$ components.}
    \label{fig:siii_lines}
\end{figure}

\begin{figure}
    \centering
    \includegraphics[width=9 cm]{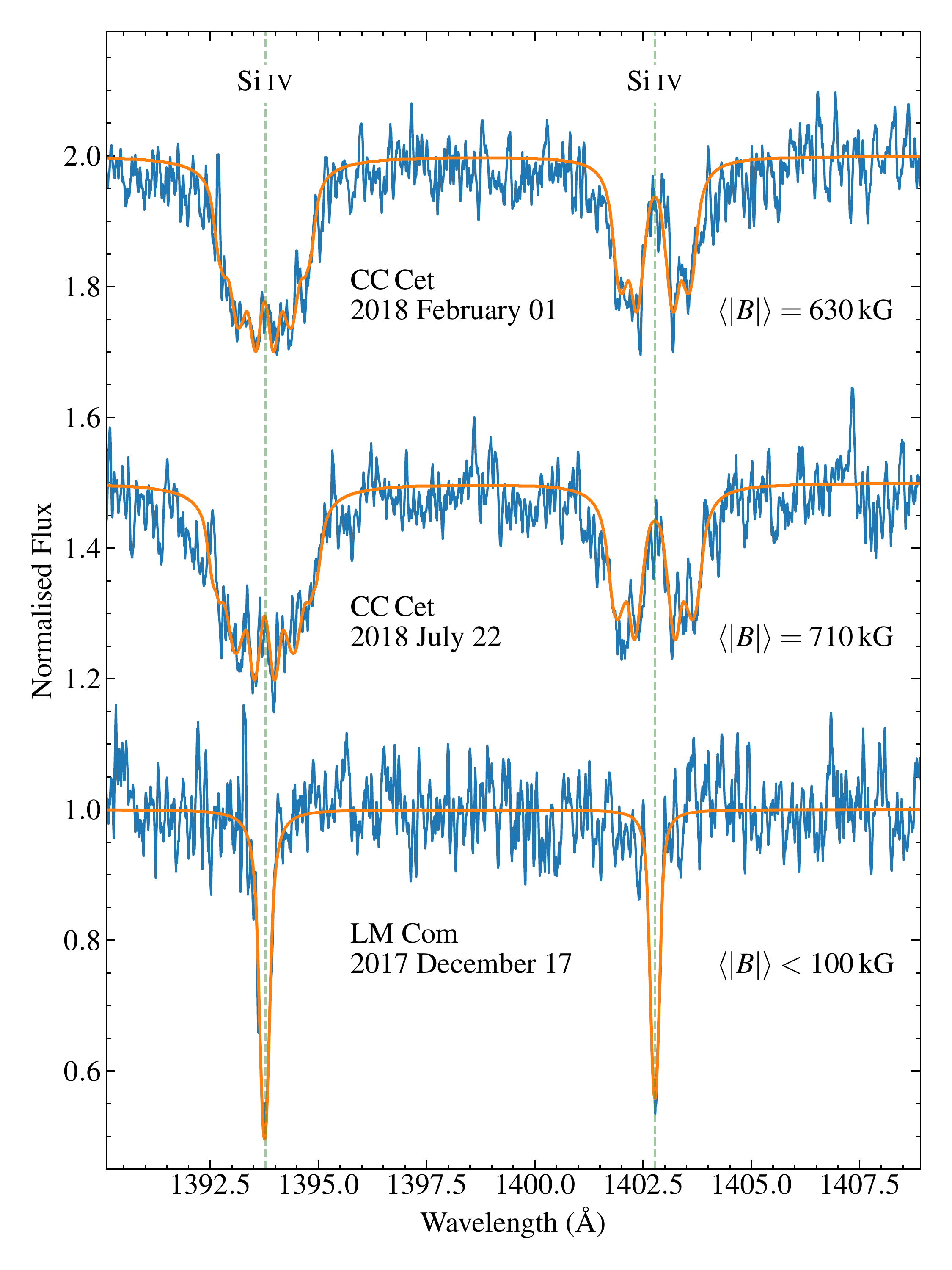}
    \caption{Silicon \ion{Si}{iv} lines in the two \textit{HST}/COS spectra of CC\,Cet, and one spectrum of LM Com, another PCEB with a similar $T_{\mathrm{eff}}$ and $\log g$. The difference between the Zeeman split lines in CC\,Cet and the non-magnetic LM\,Com is readily apparent. The orange line shows the best-fit magnetic model to the lines (in the case of LM\,Com corresponding to an upper limit on the field strength), and the dashed vertical lines show the rest wavelengths. The date and best-fit mean field modulus are given under each spectrum.}
    \label{fig:siiv_lines}
\end{figure}

\begin{figure}
    \centering
    \includegraphics[width=9 cm]{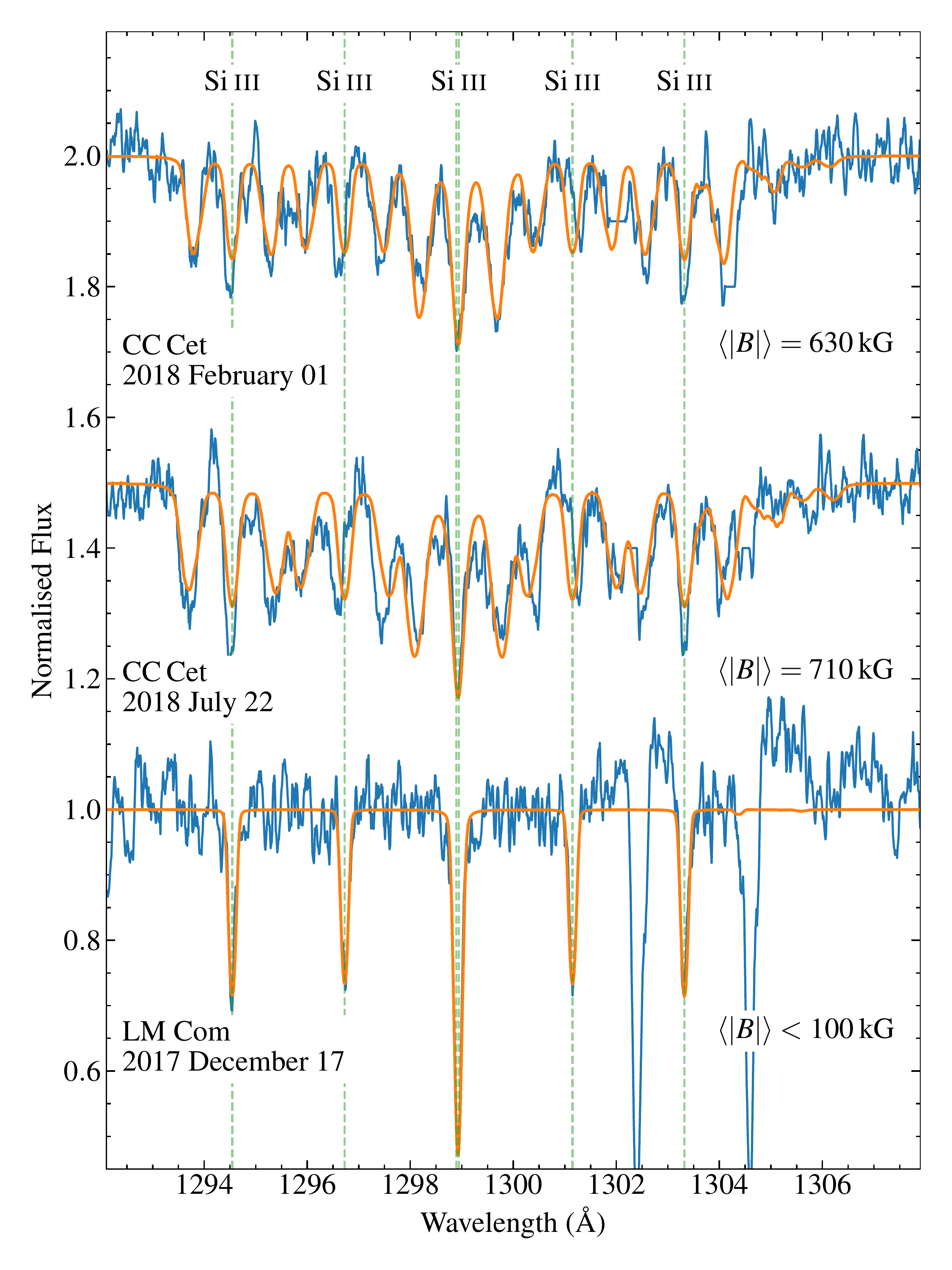}
    \caption{As Figure \ref{fig:siiv_lines} but for \ion{Si}{iii} lines around 1300\,\AA. The deep unmarked absorption features in the LM\,Com spectrum are interstellar \ion{O}{i} lines that have been removed from the spectrum of CC\,Cet. Each line in LM\,Com is replaced in the spectrum of CC\,Cet by a Zeeman triplet. Note the slight mis-match between the centroids of some of the lines in the model and the spectrum induced by the partial Paschen-Back effect.}
    \label{fig:siiii_lines}
\end{figure}

\subsubsection{Discovery of field}

Closer examination of the photospheric lines reveals that they differ significantly from those in similar COS spectra. In particular, the Si\,{\sc ii}\,1264\,\AA\ resonance line (which is slightly blended with a much weaker line in the same multiplet, but not blended with any interstellar lines because it arises from a state a few hundred cm$^{-1}$ above the ground level) shows a clear triplet structure composed of three very similar components separated by about 0.5\,\AA\ (Figure \ref{fig:siii_lines}). The structure is strongly reminiscent of the appearance of the normal Zeeman triplet produced in many spectral lines by a magnetic field of tens or hundreds of kiloGauss (kG). The appearance of this feature strongly suggests that CC\,Cet has a magnetic field.  

Furthermore, although the other strong photospheric lines do not show such obvious Zeeman splitting, they all look rather peculiar. The two \ion{Si}{iv} lines around 1400\,\AA, which normally show very similar Lorentzian profiles, instead are quite broad and different from one another (Figure \ref{fig:siiv_lines}). The six strong lines of \ion{Si}{iii} around 1300\,\AA\ appear as about 15 slightly weaker lines in the same wavelength interval (Figure \ref{fig:siiii_lines}). The \ion{C}{iii} lines at 1175\,\AA\ show only three strong lines instead of the usual five (Figure \ref{fig:ciii_ppe_obs}). As we shall show below, these effects can be produced by a magnetic field of hundreds of kG.

\subsubsection{Qualitative analysis of field}

To obtain further information about the magnetic field which appears to be present in CC\,Cet, we start by estimating the mean field modulus \bs, the value of the local field strength $|{\bf B}|$ averaged over the hemisphere visible at the time of observation. The observed splitting of the  \ion{Si}{ii}\,1264\,\AA\  line is about 0.5\,\AA\ between each of the $\sigma$ components and the central $\pi$ component. This value  may be used with the standard equation for the $\pi$ -- $\sigma$ separation $\Delta \lambda_{\rm Z}$ due to the normal Zeeman effect for an initial field strength estimate \citep[e.g.][]{Land92}:
\begin{equation}
    \label{zeeman-splitting.eqn}
\Delta \lambda_{\mathrm{Z}}(\mbox{\AA}) = 4.67\times10^{-13} z \lambda_0^2 B,
\end{equation}
where $\lambda_0$ is the unperturbed wavelength of the line, wavelengths are measured in \AA\ units, $z = 1.43$ is the mean Land{\'e} factor of this line, and $B$ is the magnetic field strength in Gauss ($10^4$\,G = 1\,Tesla). Applying this equation to the Si\,{\sc ii} line we deduce the presence of a field of $B \sim 500$\,kG.

A field of this strength is well within the regime of normal Zeeman splitting for transitions between two isolated atomic levels. However, for some of the multiplets of the light elements observed in CC\,Cet, the spacing between fine-structure levels of single spin-orbit (LS) terms is small enough that the level splitting by the Zeeman effect in a field of hundreds of kG is comparable to the unperturbed term level separation. Alternatively, one can say that the Zeeman splitting is so large that some of the outer Zeeman components approach or cross neighbouring lines of the multiplet. In this case, it is no longer correct to use the usual Zeeman splitting theory appropriate for lines formed between isolated levels. Instead, the magnetic field effect must be considered together with the fine structure splitting of the term. This case is known as the partial Paschen-Back effect, and we will discuss it below when we come to this situation.

\subsubsection{Simple dipolar magnetic field models}

Ideally, we would aim to obtain a low resolution map or model of the global structure of the magnetic field over the whole surface of the white dwarf in CC\,Cet. In principle this is possible if a series of polarised spectra are obtained through the rotation period of the star, and these are modelled collectively. However, we have only two spectra, both of mediocre S/N, without polarisation information. The stellar rotation period is fast (see below) but unknown, so we are unable to determine the net viewing angles of the observations. 

The magnetic spectrum synthesis code used for this modelling was the code {\sc zeeman.f} \citep{landstreet88-1}, which performs a forward line profile computation starting from a specified stellar atmosphere model, a specified magnetic configuration over the entire stellar surface, relevant geometric parameters such as the inclination $i$ of the rotation axis and a range of projected rotation velocities, an abundance table for relevant elements, and a specified wavelength window and line list.  The code solves the four coupled equations of polarised radiative transfer, and computes the expected emergent Stokes parameters $I$ (essentially the flux, normalised to the continuum for convenience, including emergent spectral line profiles), and the polarisation Stokes parameters $Q$ and $U$ (linear polarisation), and $V$ (circular polarisation). The output flux and polarisation profiles can then be compared to the observed spectra. One or more parameters (such as the abundance of a specific chemical element) can be iterated by the code to improve the agreement between computed and observed spectra.

Because of the limited data available for CC\,Cet we cannot fully constrain even a simple magnetic model such as a dipole field. Instead, we aimed to find a set of parameters that yield satisfactory fits to the lines in the modelled regions, and extract some general field properties from such models, such as a reasonably well-defined value of \bs, information about the extent to which different spectral lines provide concordant estimates, and an estimate of \vsini. For this modelling we restricted the field structure to a dipole, or a dipole plus a weak parallel but opposing linear octupole that has the effect of reducing the range of local $B$ variations over the surface.

\subsubsection{\ion{Si}{iv}\,1400\,\AA\ multiplet}
\label{sec: siiv}
Figure \ref{fig:siiv_lines} shows the \ion{Si}{iv} lines near 1400\,\AA, which can be fit reasonably well with a variety of simple dipole models. The two lines of this multiplet share a common lower level, and the two upper states are separated by about 460\,cm$^{-1}$, so that the lines are separated by about 9\,\AA, much more than the typical Zeeman splitting at 500\,kG. Their splitting is thus accurately described by the normal Zeeman theory, and correctly computed by {\sc zeeman.f}. A typical fit of the lines in the spectral window around 1400\,\AA\ is shown in Figure\,\ref{fig:siiv_lines}. The strange spectral line profiles of the two Si\,{\sc iv} lines in this window are easily understood as the effect of rather different Zeeman splitting patterns of the two lines. The line at 1402\,\AA\ comes from a $^2$S$_{1/2}$ to $^2$P$_{1/2}$ transition, whose Zeeman splitting pattern has the two $\pi$ components almost as far apart as the two $\sigma$ components, and thus forms effectively a Zeeman doublet, with almost no absorption at the unperturbed line centre because there are no regions of field near 0\,kG. In contrast, the six Zeeman components of the 1393\,\AA\ $^2$S$_{1/2}$ to $^2$P$_{3/2}$ transition are pretty uniformly spaced through the profile, with the most displaced and weakest $\sigma$ components at the outer edges of the line profile.  This splitting pattern leads to a roughly U-shaped overall profile. (For orientation, many simple LS-coupling Zeeman splitting patterns are shown schematically by \citealt{condon+shortley35-1}, Fig\,2$^{16}$.) The lines can be fit well with a wide variety of magnetic model parameters, provided that the mean field modulus is about 630\,kG (for the 2018~Feb spectrum) or 710\,kG (2018~Jul spectrum). The abundance of Si relative to H, $\epsilon_{\rm Si} = \log(n_{\rm Si}/n_{\rm H})$, is found from the combined best fit to the two lines to be about $-5.6 \pm 0.1$ for the 2018~Feb spectrum, and about $-5.7 \pm 0.1$ for the 2018~Jul spectrum.

\subsubsection{\ion{Si}{iii}\,1300\,\AA\ multiplet}
A second window that can be modelled fairly well in the Zeeman approximation is the set of six low excitation lines of \ion{Si}{iii} between 1294 and 1303\,\AA. These lines arise from transitions between two $^3$P terms, and a peculiarity of LS coupling leads to magnetically perturbed levels of non-zero $J$ having identical level separation with an anomalous Land{\'e} $g$-factor 1.5 (i.e. the level separation is 1.5 times larger than in the normal Zeeman separation). This splitting pattern replaces each of the single lines of the sextuplet by three lines, and the strength of the field just happens to be a value for which the added $\sigma$ components of the lines fall in between the central $\pi$ components (which lie approximately at the wavelengths of the magnetically unperturbed sextet), and replace the original lines with a forest of about 15 distinct lines, approximately equally spaced (two lines of the original sextet almost coincide in wavelength). This effect is shown in Figure\,\ref{fig:siiii_lines}, where the original lines of the multiplet (now $\pi$ components), have been marked.

In this set of lines, the separation between $\sigma$ and $\pi$ Zeeman components is comparable to the separation between lines of the sextet. Correspondingly, the magnetic splitting of some of the individual multiplet levels is comparable to the separations between both the lower and the upper term levels. In this situation the simple weak-field Zeeman splitting is no longer an accurate description; instead the splitting is making the transition to Paschen-Back splitting, and both magnetic and fine structure splitting should be computed together. Partial Paschen-Back splitting is not built into {\sc zeeman.f}, so our computations assume that the usual expression for the anomalous Zeeman effect apply. This approximation meant that the simple magnetic models that fit the Si\,{\sc iv} lines at 1400\,\AA\ did not produce good fits to the other strong lines.

\subsubsection{Partial Paschen-Back splitting}

The partial Paschen-Back effect, and how to compute its effects on the lines of a multiplet, is discussed extensively by \citet[][Sec. 3.4]{landideglinnocenti+landolfi04-1}. We obtained a {\sc fortran} program, {\sc gidi.f}, from Dr Stefano Bagnulo which was originally written by Prof. Landi Degl'Innocenti. This program solves the problem of the combined effect of magnetic and fine structure splitting of a single multiplet that is described by LS coupling. This approximation is appropriate for most light elements, including C and Si. A calculation of the splitting of the 3s3p $^3$P$^{\rm o}$ -- 3p$^2$ $^3$P transition that leads to the sextet of lines at 1300\,\AA\ indicates that the most significant effect of the partial Paschen-Back effect at 600\,kG is to shift the $\pi$ components of the 1296.72\,\AA\ and 1301.15\,\AA\ lines by about 0.12\,\AA\ bluewards and redwards respectively. Exactly this effect is clearly observed in the discrepancy between our fit using only conventional Zeeman splitting, and the observations shown in Fig.\,\ref{fig:siiii_lines}. Otherwise, the simple Zeeman effect computation of {\sc zeeman.f} provides a reasonable fit, and allows us to refine the strength of the field observed.

The value of the Si abundance  can also be determined from the model for the \ion{Si}{iii} window.  We find values of $\epsilon_{\rm Si} \approx -6.1$ to $-6.3$. Note that this abundance is about a factor of three lower than the abundance level deduced from the \ion{Si}{iv}\,1400\,\AA\ lines. The origin of this difference is not known, but may represent an overabundance of the \ion{Si}{iv}/\ion{Si}{iii} ratio produced by non-LTE effects that cannot be predicted by our LTE code, an effect of vertical stratification \citep{koesteretal14-1}, or perhaps an effect of non-uniform distribution of Si over the stellar surface, as is found on upper main sequence magnetic stars \citep[see for e.g.][]{krtickaetal19-1}.

\subsubsection{Other UV spectral lines}

\begin{figure}
    \centering
    \includegraphics[width=9 cm]{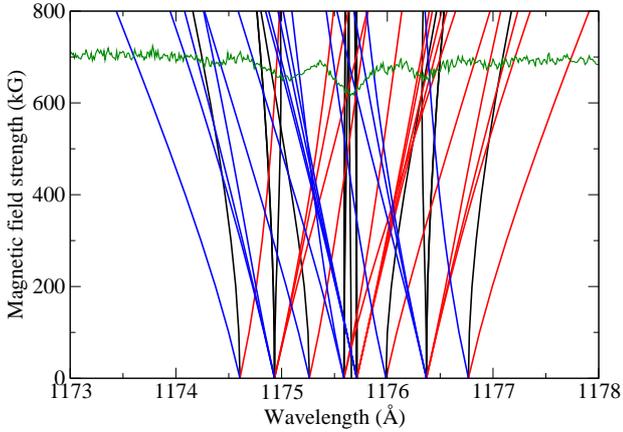}
    \caption{"Spaghetti diagram" showing wavelengths of individual magnetic subcomponents of the C\,{\sc iii} sextet at 1175\,\AA\ as a function of magnetic field strength, computed as a partial Paschen-Back case. Subcomponents are colour-coded as blue $\sigma$ (blue), $\pi$ (black), and red $\sigma$ (red). The COS flux spectrum (green) is plotted with the continuum set at about 700\,kG to show how main line features correspond with clustering of magnetic subcomponents.}
    \label{fig:ciii_ppe_obs}
\end{figure}

The  \ion{Si}{ii} line at 1265\,\AA\ (Figure \ref{fig:siii_lines}) is one of three resonance lines formed by the 3p $^2$P$^{\rm o}$ to 3d $^2$D transition. Two of the three lines nearly coincide at 1264.730 (strong) and 1265.023\,\AA\ (weak). The effect of the partial Paschen-Back effect is to shift all the components of the weaker line towards the blue, and those of the stronger line towards the red, so that radial velocities measured with the $\pi$ or $\sigma$ components of the stronger line in a field near 600\,kG will be systematically red shifted from radial velocities measured with lines unaffected by the partial Paschen-Back effect, such as those from the Si\,{\sc iv} doublet at 1400\,\AA. However, the normal Zeeman effect is not a bad approximation at 600\,kG, and models of this feature with {\sc zeeman.f} fit reasonably well assuming a field structure with $\bs \approx 550$\,kG (see Fig.\,\ref{fig:siii_lines}).

As discussed above, because the partial Paschen-Back splitting of lines is not implemented in our line synthesis code, we cannot produce an accurate model of the \ion{C}{iii} \,1175\,\AA\ multiplet. Therefore, we have computed the variation of the line splitting as a function magnetic field strength up to 800\,kG using {\sc gidi.f}. Below about 200\,kG, the splitting of the individual lines of the multiplet follows closely the prediction of the Zeeman theory. However, by about 600\,kG, the splitting is converging on the Paschen-Back limit. In fact, the computed splitting at 600\,kG quite closely resembles the observed multiplet, with five groups of line components whose centroids coincide almost exactly with the positions of the observed components, and that agree qualitatively with the relative strengths of those components (Fig.~\ref{fig:ciii_ppe_obs}).

\subsubsection{H in the optical spectrum}
\label{sec:hlines}
\begin{figure}
    \centering
    \includegraphics[width=9 cm]{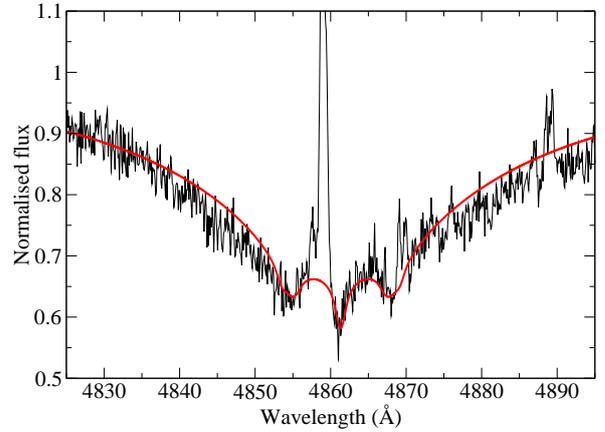}
    \caption{H$\beta$ from UVES spectrum of CC\,Cet (black) with overplotted fit to line core assuming approximately dipolar magnetic field with $\bs = 620$\,kG. The Zeeman triplet is clearly detected in spite of strong emission from the M~dwarf secondary star.}
    \label{fig:hbeta_obs_fit}
\end{figure}

The archival UVES spectra cover the entire Balmer spectrum in the visible with resolving power of about 20\,000 and S/N of roughly 15--20. Balmer lines H$\alpha$ to H$\epsilon$ are clearly visible in the spectra. We have examined the cores of these lines for evidence of the magnetic field of CC\,Cet. Because of the low S/N and the presence of emission lines from the M~dwarf companion, especially strong in H$\alpha$ and H$\beta$, the magnetic splitting of the Balmer lines by the 700\,kG field is not immediately obvious, and was missed in searches for magnetic fields in the analysis of the SPY survey \citep{napiwotzkietal20-1, koesteretal09-2}. However, if we smooth the UVES spectra slightly and shift the two spectra to the same stellar radial velocity framework, the superposed H$\beta$ lines reveal fairly clear Zeeman triplet structure, with splitting that agrees closely with that expected from the ultraviolet lines. One spectrum with a model fit is shown in Figure \ref{fig:hbeta_obs_fit}.

\subsubsection{Rotation velocity}
\label{sec:rotation}

A remarkable feature of both the \ion{Si}{iii}\,1300\,\AA\ and \ion{Si}{iv}\,1400\,\AA\ lines is that the $\pi$ components appear to be broadened beyond the point expected from the instrumental resolution. The $\pi$ components of each line are not broadened by a variation of values of the local magnetic field strength $B$ over the visible hemisphere, so the broadening may instead be due to a rapid rotation of the white dwarf. 

Measuring the cleanest $\pi$ components, we find a full width half maximum of about $\approx60$\,\kms\, and a full width at the continuum of $\approx80$\,\kms. CC\,Cet is a short-period binary system with an M4.5 main sequence star, and thus orbital motion during an exposure may be contributing to the broadening. The velocity semi-amplitude of the M\,dwarf is about 120\,\kms \citep{safferetal93-1}. The white dwarf has a mass very close to twice that of the M\,dwarf, so the velocity semi-amplitude is about 60\,\kms, with the full range being covered in about 3.5\,h. During a single 0.5\,h COS exposure, the radial velocity of the white dwarf could therefore change by as much as 25\,\kms. The lines are additionally broadened by the resolving power of the instrument ($\approx20$\,\kms) and a small wavelength spread due to the partial Paschen-Back effect ($\approx9$\kms). Summing those factors in quadrature, we find that an additional rotational broadening of $v\sin i\simeq30-40$\,\kms\ is required to match the shape and width of the observed line profile. As the contribution from the orbital velocity is an upper limit, it is unlikely that \vsini\ is actually much smaller than 40\,\kms. With this lower limit to $v_{\rm eq} \sin i$ and the radius of CC\,Cet, we can obtain an upper limit to the spin period $P = 2 \pi R/v_{\rm eq} \la 2 \pi R/v_{\rm eq} \sin i \approx 2000$\,s. Given that there are small changes in the magnetic field strength and line depth between spectra, which we infer to be due to different viewing angles of an inhomogenous surface, the rotation period cannot be much shorter than the exposure time of the observations (1865\,s) otherwise the variations in field strength would have been smeared out over multiple rotations. 

The spin period/orbital period ratio of CC\,Cet is therefore $\approx0.05$, consistent with the bulk of the intermediate polars with spin and orbital period measurements \citep{bernardinietal17-1}. The difference in the line profiles between the two spectra indicate that the magnetic field axis is inclined to the rotation axis and/or the metals are not evenly distributed across the white dwarf surface, so it may be possible to measure the rotation period from high cadence, high signal photometry or spectroscopy. No rotation signal is detected in the \textit{TESS} light curve, but the red \textit{TESS} bandpass is dominated by the M\,dwarf and may not be sensitive to subtle flux variations from the white dwarf. We attempted to produce time-series spectroscopy from the COS data using the \texttt{costools splittag}\footnote{\url{https://costools.readthedocs.io/en/latest/index.html}} routine to split the time-tag files into 30\,s bins, but the S/N became too low to reliably measure any periodic flux or absorption line variation with a 99.9\,per\,cent false alarm probability limit of $\approx0.31$\,per\,cent and $\approx0.28$\,per\,cent for the 2018\,Feb and 2018\,Jul spectra respectively.     

\subsubsection{Results of the magnetic analysis}

The following conclusions can be drawn from our magnetic modelling efforts. (1) The typical value of \bs\ on CC\,Cet is about 600--700\,kG. This value may differ by $\approx50-100$\,kG between the two COS observations. The deduced value of the field appears to be generally consistent with the splitting of all photospheric lines in the spectrum, although some observed lines (such as the sextet of lines of C\,{\sc iii} at 1175\,\AA), are split by the partial Paschen-Back effect in ways that our modelling code cannot reproduce accurately. (2) The fact that the centre of the 1402\,\AA\ line reaches almost to the continuum shows that there are no large regions of field strength close to zero on the visible surface. These facts are consistent with, but do not strongly require, a roughly dipole-like field, similar to those found in some other low field magnetic white dwarfs that have been modelled in detail \citep[e.g. WD\,2047+372: see][]{landstreetetal17-1}. (3) CC\,Cet has $v \sin i \approx 40$\,\kms, and therefore has a rotation period \la 2000\,s, but not much shorter than that due to the differences seen between the two 1865\,s exposures. 

\subsection{X-ray Analysis}
\label{sec:x-rays}
Since there were too few X-ray events to perform any sort of spectral analysis, in order to understand the source X-ray luminosity that might have given rise to a weak signal we used the {\tt PIMMS} software\footnote{\url{https://heasarc.gsfc.nasa.gov/docs/software/tools/pimms.html}} version 4.11 to convert between pn and MOS count rates and incident X-ray flux. This was done for the APEC optically-thin plasma radiative loss model \citep{fosteretal12-1} for the solar abundances of \citet{asplundetal09-1} and an intervening hydrogen column density of $2\times 10^{20}$~cm$^{-2}$. The latter was estimated from the distance of 121\,pc  \citep{gaia18-1} and interpolation within the neutral hydrogen column density compilations of \citet{Linsky.etal:19} and \citet{Gudennavar.etal:12}.  

\begin{figure}
    \centering
    \includegraphics[width=0.47\textwidth]{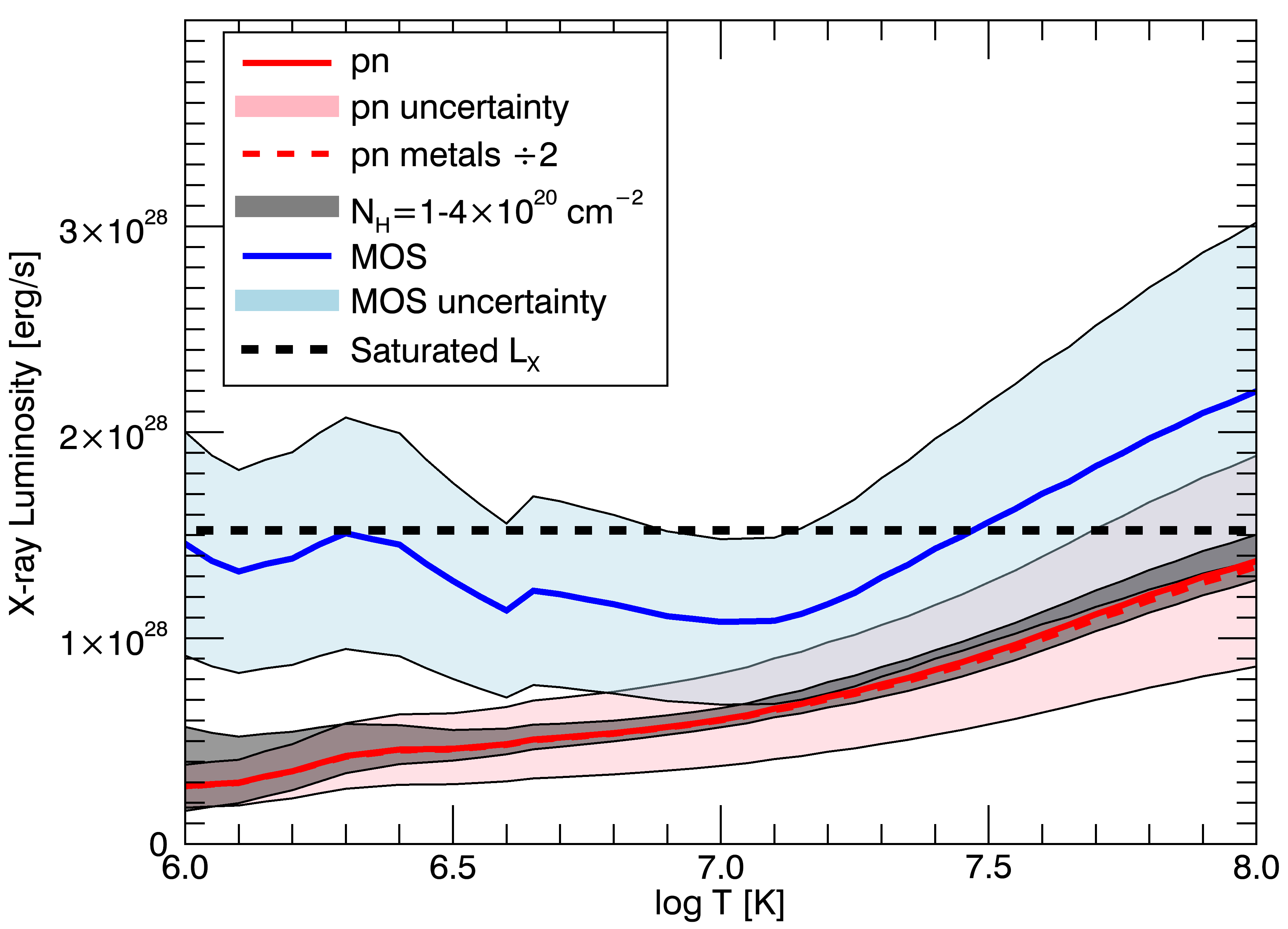}
    \caption{The X-ray luminosities obtained from {\it XMM-Newton} pn and MOS count rates assuming the X-ray signal originates in a hot isothermal  collision-dominated  optically-thin plasma. The derived luminosities are shown as a function of isothermal plasma temperature. Uncertainty ranges due to Poisson statistics in the observed count rates, and to uncertainties in the intervening interstellar medium absorbing column are indicated by shaded regions. The sensitivity to plasma metallicity is very low and indicated for the pn case only. The saturated coronal X-ray emission level for a star of the spectral type of the CC\,Cet secondary is also illustrated.}
    \label{f:xmmlx}
\end{figure}

The resulting X-ray luminosities in the 0.3--10\,keV band corresponding to the  observed pn and MOS count rates are illustrated as a function of isothermal plasma temperature in Figure~\ref{f:xmmlx}. The  luminosities for a typical active stellar coronal temperature of $10^7$\,K are also listed in Table~\ref{t:xmm}. Sensitivity of the results to the adopted absorbing column was examined by computing the analogous luminosities for values of $N_\mathrm{H}$ lower and high by a factor of two, while sensitivity to metallicity was checked by computing luminosities for metallicity reduced by a factor of two. These cases are also illustrated in Figure~\ref{f:xmmlx}.  By far the largest uncertainty is in the estimate of the X-ray count rates. Luminosities derived from the MOS data appear larger than those from pn, but again the uncertainties overlap for most of the temperature range expected for active stellar coronae.

Also shown in Figure~\ref{f:xmmlx} is the canonical X-ray saturation luminosity, $L_\mathrm{X}/L_\mathrm{bol}=10^{-3}$, corresponding to an M4.5~V star with a luminosity of $0.0040\mathrm{L}_\odot$ \citep{pecaut+mamajek13-1}.  Our very tentative detection of CC\,Cet then lies essentially at the saturation limit.  This is expected since the orbital period of the binary, and presumably the rotation period of the secondary M~dwarf assuming tidal synchronization, is well into the saturated regime that sets in at rotation periods shorter than approximately 20 days for a mid-M~dwarf \citep[e.g.][]{Wright.etal:11}.

\section{Discussion}
\label{sec:dis}

\begin{figure}
    \centering
    \includegraphics[width=8 cm]{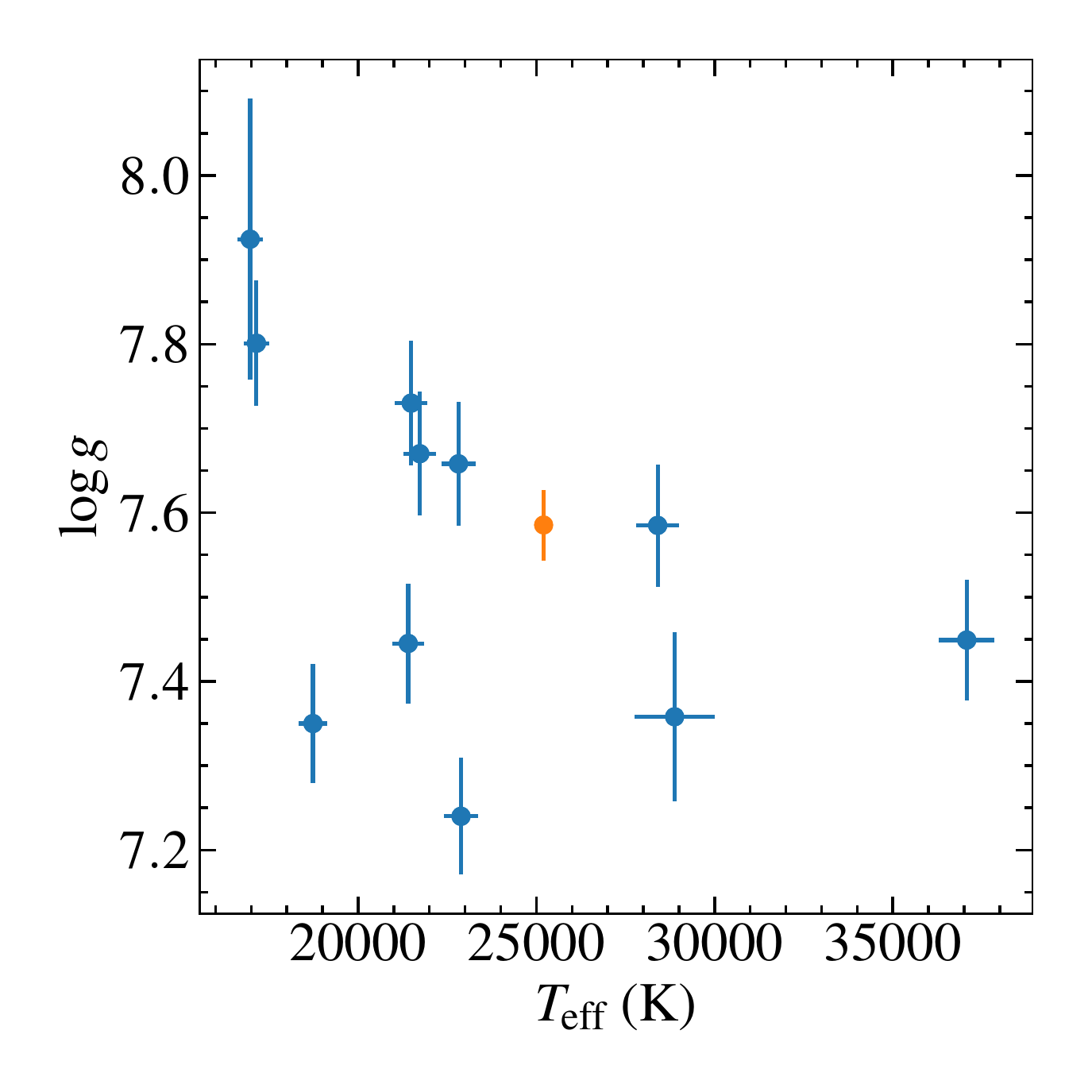}
    \caption{Atmospheric parameters of the white dwarfs in PCEBs observed with \textit{HST}/COS, demonstrating that CC\,Cet (orange) is not an outlier in either $T_{\mathrm{eff}}$ or $\log g$.}
    \label{fig:tefflogg}
\end{figure}

\begin{figure}
    \centering
    \includegraphics[width=8 cm]{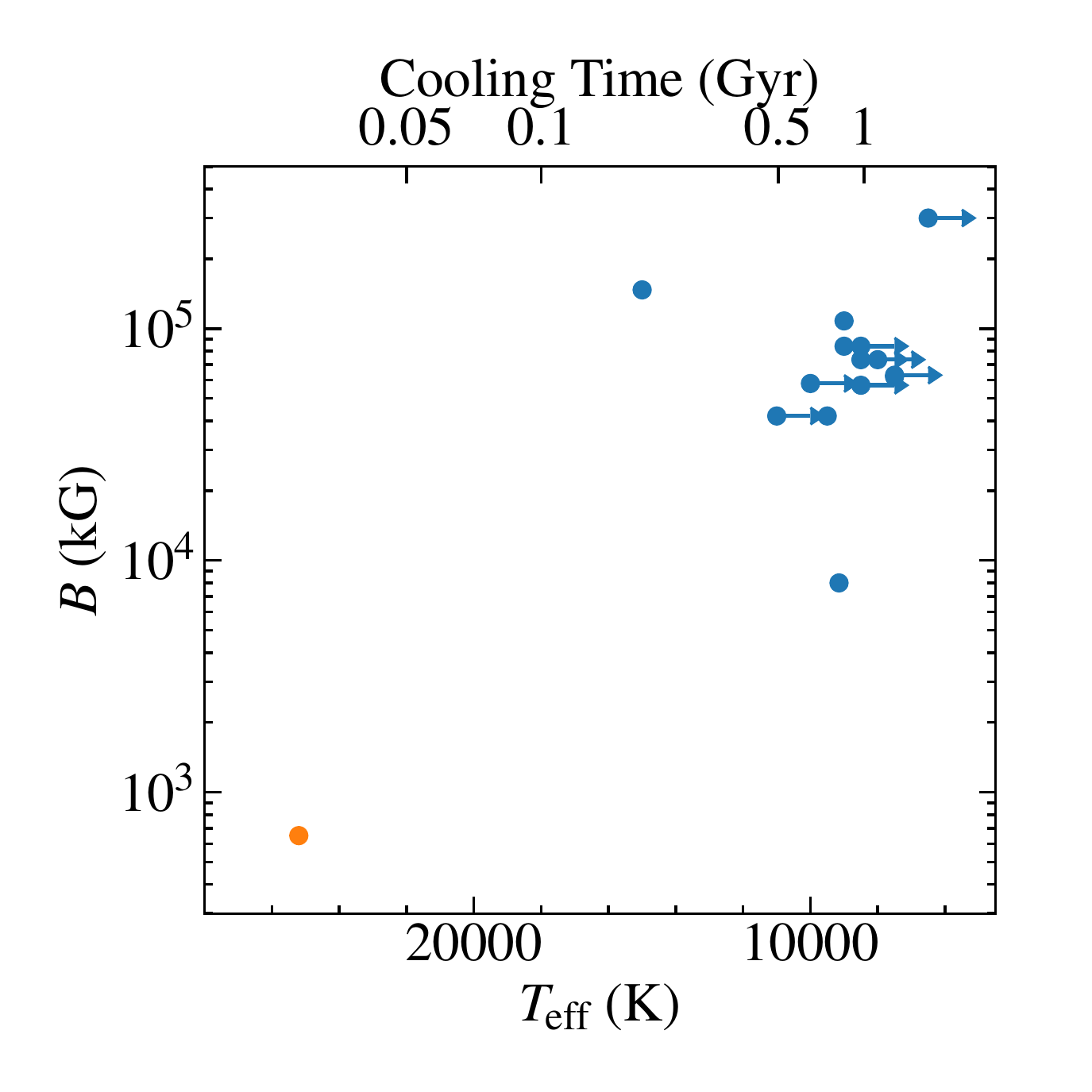}
    \caption{Temperatures and magnetic field strengths of known magnetic white dwarfs in PCEBs compiled by \citet{ferrarioetal15-1} and \citet{parsonsetal21-1} (blue) compared with CC\,Cet (orange). An approximate cooling time scale is given on the top axis. CC\,Cet is clearly an extreme outlier in both \Teff\ and $B$.}
    \label{fig:teffB}
\end{figure}

\subsection{CC Cet in context}
Along with CC\,Cet, program 15189 observed 9 other PCEBs containing a white dwarf and an M~dwarf with the same COS setup. Archival COS G130M spectra of three additional PCEBs exhibiting metal absorption lines are also available (Programs 12169,12474,12869). Although a full analysis of these observations will be left for a future publication, we can use them to estimate an occurrence rate for magnetic PCEBs. All the other 12 white dwarf spectra show metal absorption lines, with no detectable magnetic field (\bs\ $\lesssim100$\,kG). Although the PCEBs were selected for practicality of observation, rather than to provide an unbiased sample, a comparison of the white dwarf atmospheric parameters (Figure \ref{fig:tefflogg}) demonstrates that CC\,Cet is typical of the sample. Using the \texttt{astropy.stats.binned\_binom\_proportion} function we calculate an occurrence rate for magnetic white dwarfs of $7.7^{+14}_{-2.5}$\,per\,cent, where the uncertainties are the $1\sigma$ confidence boundaries. This stands in stark contrast to the 0/1735 ($<0.11$\,per\,cent) and 2/1200 ($0.17^{+0.22}_{-0.05}$\,per\,cent) rates found by \citet{liebertetal15-1} and \citep{silvestrietal07-1} respectively. The reason for the discrepancy is probably the different waveband used here to the two previous studies: the ultraviolet wavelength range is riddled with sharp metal absorption lines that are sensitive tracers of even relatively low fields, which will go unrecognised in low-resolution optical spectra such as those obtained by the Sloan Digital Sky Survey, and are difficult to detect even in high-resolution spectra of the Balmer lines (see Section\,\ref{sec:hlines}, Figure \ref{fig:hbeta_obs_fit}). We note that the fraction of intermediate polars among the CVs in the nearly complete 150\,pc sample of \citet{palaetal20-1} is $3/42$ or $7.1^{+6.6}_{-2.2}$\,per\,cent, i.e. consistent with the incidence of magnetic white dwarfs in the COS PCEB sample derived above.

Figure \ref{fig:teffB} compares CC\,Cet with the effective temperatures and magnetic field strengths of the magnetic PCEBs compiled by \citet{ferrarioetal15-1} and the latest discoveries by \citet{parsonsetal21-1}. CC\,Cet is clearly an extreme outlier, being both much hotter (and therefore younger) than the rest of the sample and having a much weaker magnetic field strength, even in comparison to the other confirmed pre-intermediate polar SDSS\,J030308.35+005444.1. CC\,Cet is an outlier by $\approx6\sigma$ in both $\log (B)$ and \Teff (where the stars that only have upper limits on \Teff\ were treated as being at the upper limit). 

\subsection{Formation and evolution}

\subsubsection{Formation}
Invoking a common origin for the magnetic white dwarfs in Figure \ref{fig:teffB} would imply a large and thus-far undetected population filling the parameter space in between CC\,Cet and the rest of the sample. We speculate that it is instead more likely that the CC\,Cet magnetic field was formed via a different pathway to those of the previously known pre-polars. Such a pathway must consistently explain the low but non-zero occurrence rate of magnetic white dwarfs in PCEBs, as well as the high rotation rate and low mass of the CC\,Cet white dwarf.
Here we discuss the various proposed formation scenarios for magnetic white dwarfs and their applicability to CC\,Cet:

\textit{Spin up from accretion:} The crystallization/spin up model \citep{iserneral17-1, schreiberetal21-1} successfully reproduces the existing examples of cool, high-field pre-polars. However, this model require the system to have undergone a period of mass transfer via Roche lobe overflow. As we will discuss in Section \ref{sec:future}, the CC\,Cet secondary will not fill its Roche lobe and start mass transfer for many Gyr. Furthermore, the system is too young for the white dwarf core to have adequately crystallized. Our measured mass for the CC\,Cet white dwarf, $0.441\pm0.008$\,\Msun, indicates that it may be a He-core white dwarf \citep{driebeetal98-1, althausetal13-1}, and thus may lack altogether the core chemical stratification necessary to generate a dynamo via convection. 

\textit{Fossil Field:} In the fossil field model, the progenitor of the white dwarf is an Ap/Bp star with the magnetic field in place before the common envelope. Whilst we cannot conclusively rule this pathway out, there are several issues. Firstly, the formation of the Ap/Bp star is thought to be due to a main sequence stellar merger \citep{braithwaite+spruit04-1}, so the average masses of the progenitor stars, and by extension their white dwarf remnants, should be higher than non-magnetic objects \citep{ferrarioetal20-1}. CC\,Cet's mass is instead lower than the average masses of both PCEBs and CVs \citep{zorotovicetal11-1}. The fossil field pathway also provides no explanation for the high rotation rate of the CC\,Cet white dwarf.

\textit{Common envelope dynamo:} \citet{briggsetal18-1} proposed a model whereby a weak initial field is wound up by differential rotation between the two stellar components in a common envelope. Whilst \citet{bellonietal20-1} found that this model was unable to predict the observed magnetic white dwarf population, it may still be relevant for rare objects such as CC\,Cet. The fact that the CC\,Cet white dwarf is already rotating with a period similar to those seen in intermediate polars may also hint at unusual dynamical interactions in the common envelope. Using Equation 1 of \citet{briggsetal18-1} and assuming that the current orbital period of CC\,Cet is close to the initial post-common envelope orbital period, we find a predicted field strength of $\approx20 $\,MG, roughly 30 times the measured value. The common envelope dynamo model as it stands cannot therefore fully explain the CC\,Cet system, but could still, at least conceptually, provide a formation scenario for the magnetic white dwarfs in CC\,Cet and the intermediate polars. However, any future extension of this model will also have to explain why \textit{most} systems emerging from a common envelope are not magnetic (Fig.\,\ref{fig:tefflogg}).

\subsubsection{Future of CC\,Cet}
\label{sec:future}

The future of CC\,Cet was first modelled by \citet{schreiber+gaensicke03-1}. They found that the companion will fill its Roche-lobe and start mass transfer onto the white dwarf, becoming a cataclysmic variable once the system has evolved down to an orbital period of $P_\mathrm{orb}\simeq2$\,h in $\simeq18$\,Gyr. 

Here we improve on those calculations using the state-of-the-art stellar evolution code Modules for Experiments in Stellar Astrophysics  \citep[{\tt MESA} v.12778,][]{paxtonetal11-1, paxtonetal13-1, paxtonetal15-1, paxtonetal18-1, paxtonetal19-1}. The standard model of evolution for CVs states that the orbital period is reduced via angular momentum losses driven by gravitational wave radiation \citep{paczynski67-1} and magnetic braking generated by the secondary (donor) star \citep{verbunt+zwaan81-1, rappaportetal83-1, mestel+spruit87-1, kawaler88-1, andronovetal03-1}. The low mass of the donor star in CC\,Cet indicates that it is fully convective, so magnetic braking is not active and the reduction of CC\,Cet's orbit is driven only by the angular momentum loss via emission of gravitational waves.

For our simulations we adopt an initial orbital period of 0.287\,d and treat the white dwarf as a point source with a mass 0.44\,\Msun (Table \ref{tab:characteristics}). We assumed that the white dwarf retains none of the accreted mass (\textsc{mass\_transfer\_beta} = 1.0, where the typical assumption is that the accreted mass is ejected in classical nova eruptions) and adopted the $M_{\rm donor}=0.18\pm0.05$\,\Msun\ secondary mass from \citet{safferetal93-1}. We ran the simulations with two different prescriptions for the gravitational wave radiation: (1) the classical prescription dictated by Einstein's quadrupole formula \citep{paczynski67-1} and (2) the calibrated version, which reproduces the observed masses and radii of the donors in cataclysmic variables \citep{kniggeetal11-1}. The simulations were ended when the mass of the donor reached the brown dwarf mass limit ($\mbox{\textsc{star\_mass\_min\_limit}}=0.08\,\Msun$).

\begin{figure}
    \centering
    \includegraphics[width=8cm]{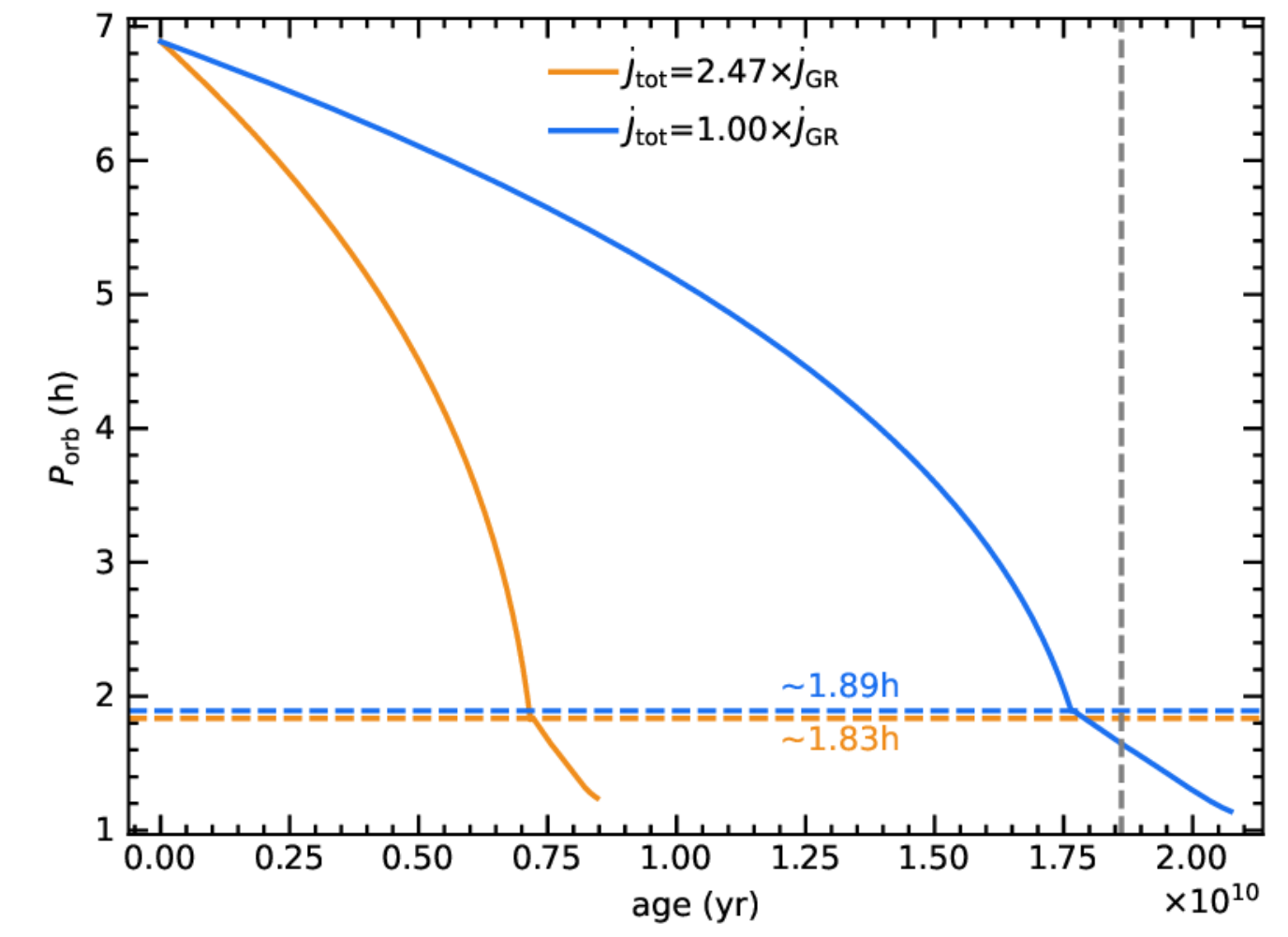}
    \caption{MESA simulations adopting angular momentum loss via gravitational wave radiation using the classical prescription \citep[][blue]{paczynski67-1} and the calibrated version \citep[][ orange]{kniggeetal11-1}. Our simulations show that CC\,Cet will start mass transfer from the secondary onto the white dwarf (i.e. become a cataclysmic variable) in the next 17.7\,Gyr and 7.2\,Gyr, for the classical and the calibrated versions of gravitational radiation, respectively. The vertical dashed line shows the calculations performed by \citet{schreiber+gaensicke03-1} using the classical prescription, which is slightly longer than in our simulations. The system will have an orbital period shorter than 2\,hr when the accretion starts (dashed lines.) }
    \label{fig:MESA}
\end{figure}

The results of our simulations confirm that the final fate of CC\,Cet is to become an intermediate polar. The system will start to transfer mass at a low rate of $\simeq 3.5\times10^{-11}$\,\Msun/yr ($\simeq 10^{-10}$\,\Msun/yr) in roughly 17.7\,Gyr (7.2\,Gyr) for the classical (calibrated) version of gravitational radiation. The system will have an orbital period of $\approx 1.89$ ($\approx 1.83$)\,h when the accretion starts (Figure \ref{fig:MESA}). The accretion rate will slightly decrease while the system continues shrinking, until the mass of the donor reaches a mass of 0.08\,\Msun\ at an orbital period of $\simeq1.14$\,h ($\simeq1.24$\,h). 

\citet{nortonetal04-1} modelled the magnetic moment of intermediate polars given their measured  spin period to orbital period ratios. Given that we know the magnetic moment of CC\,Cet ($\mu_1 = B\,R^3 \approx1.35\times10^{33}$\,G\,cm$^{-3}$), we can use their Figure~2 to estimate the white dwarf spin period when it reaches the orbital period minimums calculated above. We find equilibrium spin periods of $\simeq1.7$\,min ($\simeq1.9$\,min) for orbital periods of $\simeq1.14$\,h ($\simeq1.24$\,h),  comparable to the spin period of V455\,And, an intermediate polar near the orbital period minimum \citep{araujo-betancoretal05-1, bloemenetal13-1}.

\subsection{M~dwarf wind}
\label{s:wind}

As mass-transfer by Roche lobe overflow has not yet begun in the CC\,Cet system, the origin of the metals in the white dwarf atmosphere is most likely the stellar wind of the M~dwarf companion. \citet{debes06-1} has demonstrated that such wind accreting systems can be used to quantify the wind mass-loss rates of M~dwarfs, which is otherwise extremely difficult to measure (see \citealt{woodetal21-1} for a summary of the current state of the art). \citet{debes06-1} assumed that the non-magnetic white dwarfs in their sample accreted via the Bondi-Hoyle process \citep{bondi+hoyle44-1}, where the white dwarf gathers a fraction of the stellar wind in proportion to its mass and the wind velocity. \citet{webbink+wickramasinghe05-1} demonstrated that for binaries containing white dwarfs with high (10s of MG, i.e pre-polars) magnetic fields, the energy density of the magnetic field is much stronger than that of the wind down to the surface of the secondary, and thus the white dwarf accretes all of the wind emitted. CC\,Cet is an intermediate case, where the magnetic field is not strong enough to accrete all of the wind but nevertheless gathers wind from a wider radius than the Bondi-Hoyle process. Following the model from Section 6 of \citet{webbink+wickramasinghe05-1}, the wind will be accreted inside of a critical radius $r_{\mathrm{crit}}$ from the white dwarf where the energy density of the magnetic field  exceeds that of the wind:

\newcommand{\D}{\displaystyle}
\begin{equation}
    \label{eqn: rcrit}
\frac{\left(B_0 \left(\frac{\D r_{\mathrm{wd}}}{\D r_{\mathrm{crit}}}\right)^3\right)^2}{8\pi} = u_0\left(\frac{r_{\mathrm{md}}}{a-r_{\mathrm{crit}}}\right)^2 
\end{equation}

Where $B_0$ is the strength of the magnetic field at the white dwarf surface, $u_0$ the energy density of the wind at the secondary surface, $a$ the orbital separation, and $r_{\mathrm{wd}}$ and $r_{\mathrm{md}}$ the radii of the primary and secondary, respectively. Taking the average of our two magnetic field measurements and adopting a secondary radius of $r_{\mathrm{md}}=0.243$\,\Rsun\ from the latest version of the tables from \citet{pecaut+mamajek13-1}\footnote{\url{http://www.pas.rochester.edu/~emamajek/EEM_dwarf_UBVIJHK_colors_Teff.txt}}, we numerically solve equation \ref{eqn: rcrit} to find a critical radius of $\approx$ 1.2\,\Rsun, 75\,per\,cent of the orbital radius and roughly twice the Bondi radius. Assuming (incorrectly, but close enough for our purposes) that the M\,dwarf emits wind uniformly in all directions, the accretion rate of wind siphoned onto the white dwarf $\dot{M}_{\mathrm{wd}}$ is:
\begin{equation}
    \label{eqn: wdfrac}
\dot{M}_{\mathrm{wd}} = \dot{M}_{\mathrm{wind}} \left(\frac{r_{\mathrm{crit}}}{a}\right)^2      
\end{equation}
To calculate $\dot{M}_{\mathrm{wd}}$, we take the average of our measured Si abundances and calculate an accretion rate assuming that the atmosphere is in accretion diffusion equilibrium \citep{koester10-1}, using the diffusion timescales from the Montreal White Dwarf Database\footnote{\url{http://www.montrealwhitedwarfdatabase.org/evolution.html}}. We then assumed that the wind has Solar abundances, so scaled the Si accretion rate to the total accretion rate via the Si mass fraction measured by \citet{asplundetal09-1}, arriving at an accretion rate onto the white dwarf of $\approx7.3\times10^9$\,g\,s$^{-1}$. Via equation \ref{eqn: wdfrac}, we therefore infer a wind mass loss rate $\dot{M}_{\mathrm{wind}}$ of $\approx1.3\times10^{10}$\,g\,s$^{-1}$ or $\approx2\times10^{-16}$\,\Msun\,yr$^{-1}$. Repeating the calculation with only Bondi-Hoyle accretion in effect results in an inferred wind rate approximately an order of magnitude higher. These values are comparable to the wind mass loss rates found by \citep{debes06-1}.

X-ray observations may also be used to place constraints on the wind mass loss rate. Although the weak X-ray detection at CC\,Cet can be satisfactorily accounted for by coronal emission (Section \ref{sec:x-rays}), it is still possible that some fraction of the signal could be due to accretion onto the white dwarf. The maximum X-ray luminosity that can be extracted from accretion by conversion of gravitational potential energy to X-rays is $L_\mathrm{X}\leq GM_\mathrm{wd}\dot{M}/R_\mathrm{wd}$. For an X-ray luminosity of $L_\mathrm{X}\sim 10^{28}$\,erg\,s$^{-1}$, and the white dwarf mass and radius derived in Sect.~\ref{sec:wdparams}, the corresponding maximum mass accretion rate is $\sim 3\times 10^{-15} \Msun$\,yr$^{-1}$. This is of the same order of magnitude as estimates of the mass loss rate for the moderately active mid-M dwarf Proxima Centuri \citep[e.g.][]{wood18-1,Wargelin.Drake:02}. Since the white dwarf is unlikely to accrete a majority of the stellar wind, a more reasonable accretion efficiency of $\sim 10$\% would imply a mass loss rate of $3\times 10^{-14} M_\odot$~yr$^{-1}$, which is similar to that of the Sun. As this is two orders of magnitude higher than the mass loss rate estimated from the photospheric metal absorption lines, we retain the former value as the final wind mass loss rate. 

\section{Conclusions}
\label{sec:conc}
We have detected a 600--700\,kG magnetic field on the white dwarf component of the detached PCEB CC\,Cet, classifying it as a pre-intermediate polar. Analysis of \textit{HST} COS spectra demonstrates that the white dwarf is accreting the stellar wind of its M4.5--5 companion inhomogenously over its surface, and that the axis of the magnetic field is likely offset from the spin axis of the white dwarf. The white dwarf has a relatively low mass of $0.441\pm0.008$\,\Msun, and is rotating with a period of $\la 2000$\,s, which is much faster than the $6.88233\pm0.00045$\,h binary orbital period and consistent with the high spin period to orbital period ratios of intermediate polars.

CC\,Cet is by far the youngest and has the weakest field of all known magnetic white dwarfs in detached PCEBs, being a 6\,$\sigma$ outlier in both \Teff\ and $B$. Using MESA stellar evolution models, we show that the secondary star will not start mass transfer for at least 7\,Gyr, and is so far from filling its Roche lobe that it cannot have undergone a period of mass transfer in the past, unlike the rest of the known pre-polars. The CC\,Cet magnetic field therefore cannot have formed via mass transfer, ruling out the formation pathway proposed by \citet{schreiberetal21-1} for the pre-polars. The CC\,Cet magnetic field must instead have formed either before or during the common envelope phase, although neither the fossil field or common envelope interaction models provide a complete explanation of the observed white dwarf properties. 

The occurrence rate of magnetic to non-magnetic white dwarfs in the available sample of PCEBs with ultraviolet spectra is consistent with the fraction of intermediate polars among CVs in the 150\,pc sample. However, with only one known example in a relatively small sample, we can place only limited constraints on both the true occurrence rate and formation pathway of CC\,Cet-like
systems. The online catalogue of white dwarf plus main sequence binaries compiled by \citet{rebassa-mansergasetal12-1}\footnote{\url{https://www.sdss-wdms.org/}} contains 66 objects with \textit{GALEX} FUV magnitudes $<16$, roughly the limit where ultraviolet spectroscopy of similar precision to CC\,Cet ($FUV=14.03$\,mag) can be obtained. The periods of most of these targets are currently unconstrained, so in many cases the binary separation may be too large for stellar wind accretion to form the strong absorption lines required to search for magnetic fields. If periods for these systems can be measured, and are favourable for wind accretion, then a factor 2--3 increase in the number of PCEBs with ultraviolet spectra may be achievable, further constraining the occurrence rate and/or discovering new examples of magnetic systems.       

\section*{Acknowledgements}
We thank the anonymous referee for a pleasant and constructive review. BTG was supported by a Leverhulme Research Fellowship and the UK STFC grant ST/T000406/1. JDL acknowledges support from the Natural Sciences and Engineering Research Council of Canada (NSERC), funding reference number 6377--2016. OT was supported by a Leverhulme Trust Research Project Grant and FONDECYT project 321038. Support for this work was in part provided by NASA {\em TESS} Cycle 2 Grant 80NSSC20K0592.

We acknowledge the contribution to this work of the late Professor Egidio Landi Degl'Innocenti, whose computer program {\sc gidi.f} was supplied to us by Dr. S. Bagnulo.

Based on observations made with the NASA/ESA Hubble Space Telescope, obtained from the Data Archive at the Space Telescope Science Institute, which is operated by the Association of Universities for Research in Astronomy, Inc., under NASA contract NAS 5-26555, and observations obtained with \textit{XMM-Newton}, an ESA science mission with instruments and contributions directly funded by ESA Member States and NASA. These observations are associated with program \# 15189. We thank the {\em HST} and {\em XMM} support teams for their work arranging two simultaneous observations of CC\,Cet.

This paper includes data collected by the TESS mission. Funding for the TESS mission is provided by the NASA's Science Mission Directorate.

Based on data obtained from the ESO Science Archive Facility.

\section{Data Availability}
All of the observational data used in this paper is publicly available and can be retrieved from the relevant online archives (see Table \ref{tab:hst_obs}).




\bibliographystyle{mnras}
\bibliography{aamnem99,aamorebib,aabib,boris}




\appendix
\section{Supplementary Material}

\begin{table*}
\centering
\caption{Summary of spectroscopic observations of CC\,Cet. Dataset numbers are given for retrieval from MAST (\url{https://archive.stsci.edu/hst/}), the XMM-Newton Science Archive (\url{http://nxsa.esac.esa.int/nxsa-web/\#home}) and or the ESO Archive Science Portal (\url{http://archive.eso.org/scienceportal/home}).}  
\begin{tabular}{lcccccc}\\
\hline
Date & Instrument & Grating &  Central Wavelength (\AA) & Start Time (UT) & Total Exposure Time (s) & Dataset \\
\hline 
\textit{HST} & & & & & & \\
2018-02-02 & COS &	G130M &	1291 & 23:50:53 & 	1865 &	LDLC01010 \\
2018-07-22 & 	COS &	G130M &	1291	& 05:40:51 & 1865 &	LDLC51010\\
\textit{XMM} & & & & & & \\
2018-02-01 & -- & -- & -- & 22:48:07 & 9500 & 0810230101\\
2018-02-01 & -- & -- & -- & 22:48:07 & 16800 & 0810231301\\
VLT  & & & & & & \\
2001-02-07 & UVES & -- & 5635  & 01:27:42.810 & 600 & ADP.2020-09-01T15:53:01.123\\
2001-02-07 & UVES & -- & 3922 & 01:27:44.233 &600 & ADP.2020-09-01T15:53:01.174\\
2001-02-08 & UVES & -- & 3922 & 01:36:22.407 &600 & ADP.2020-09-01T15:54:42.325\\
2001-02-08 & UVES & -- & 5635 &01:36:21.194 & 600 & ADP.2020-09-01T15:54:42.356 \\

\hline

\hline
\end{tabular}
\label{tab:hst_obs}
\end{table*}

\begin{table}
    \centering
    \caption{Interstellar absorption lines identified in the COS spectrum of CC-Cet. These lines are narrower than the white dwarf's photospheric absorption lines because they are not affected by the rotational broadening of the white dwarf.}
    \label{tab:ISlines}
    \begin{tabular}{c l}
        \hline \hline
        Ion & wavelength (\AA; vacuum) \\
        \hline
        \Ion{C}{ii}   & 1334.532, 1335.703\\
        \Ion{N}{i}    & 1134.165, 1134.415, 1134.980, 1199.549, 1200.224, 1200.711\\
        \Ion{O}{ii}   & 1302.168\\
        \Ion{Si}{ii}  & 1190.416, 1193.289, 1260.421, 1304.372\\
        \Ion{S}{ii}   & 1250.586, 1253.812, 1259.520\\
        \hline
    \end{tabular}
\end{table}

\begin{figure}
    \centering
    \includegraphics[width=\columnwidth]{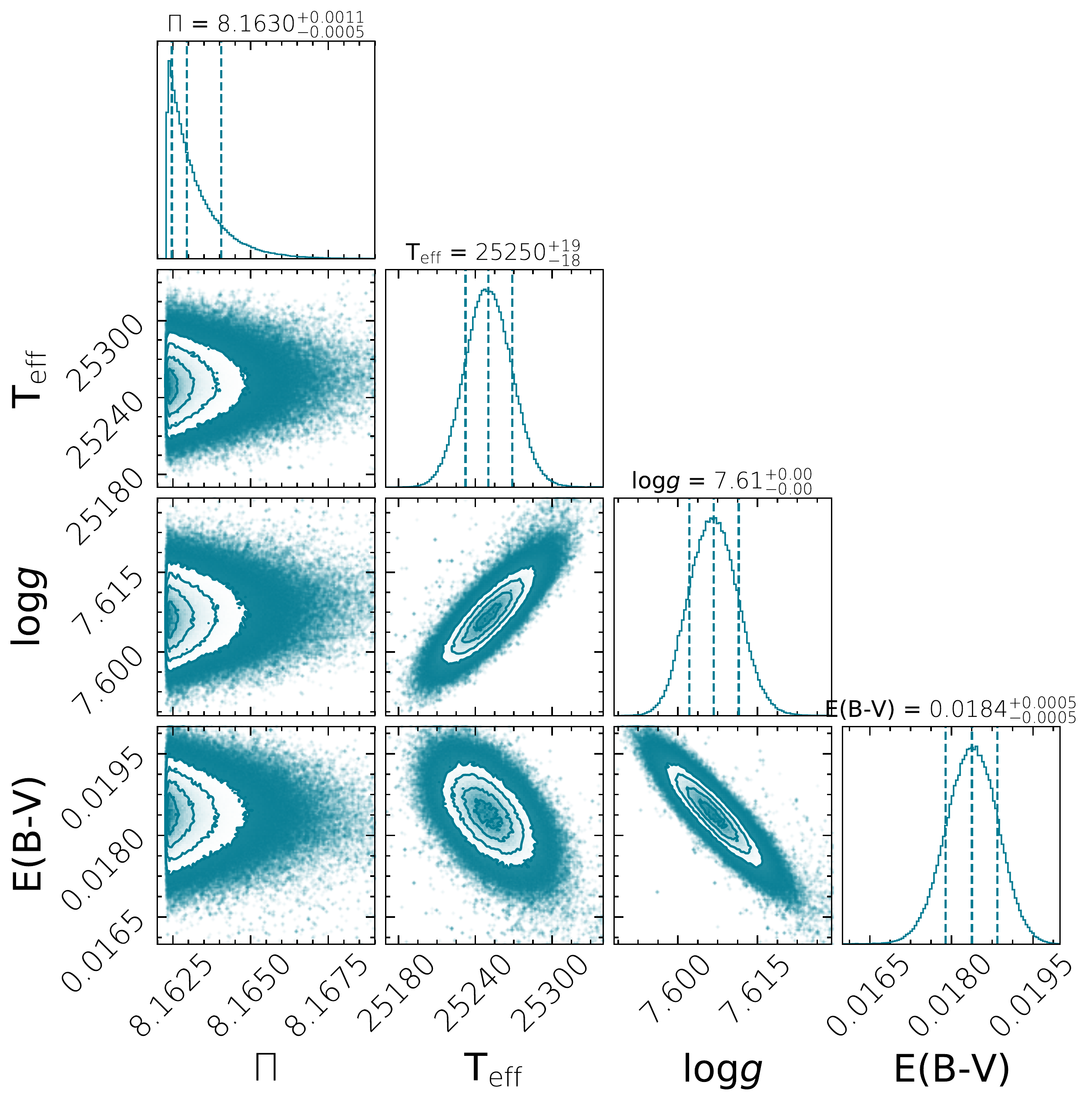}
    \caption{Results of the MCMC fit to the white dwarf atmospheric parameters plotted with {\tt corner.py} \citep{corner}.}
    \label{fig:mcmc}
\end{figure}


\bsp	
\label{lastpage}
\end{document}